\documentclass[conference,compsoc]{IEEEtran}

\usepackage{url}
\usepackage{color}
\usepackage{xcolor}
\usepackage{booktabs}
\usepackage{multirow}
\usepackage{tabularx}
\usepackage{graphicx}
\usepackage{xspace}

\ifCLASSOPTIONcompsoc
  \usepackage[nocompress]{cite}
\else
  \usepackage{cite}
\fi

\ifCLASSINFOpdf
 
\else
 
\fi

\usepackage{xcolor,listings}
\definecolor{codegreen}{RGB}{0,125,0}
\definecolor{codepurple}{RGB}{170,0,150}
\lstdefinestyle{cppcode}{
  language=C++,
  basicstyle=\scriptsize\ttfamily,
  keywordstyle=\color{blue},
  morekeywords={size_t},
  deletekeywords={if,else},
  morekeywords=[2]{if,else},
  keywordstyle=[2]\color{codepurple},
  commentstyle=\color{codegreen},
  frame=single,
  numbers=left,
  numbersep=6pt,
  columns=fullflexible,
  breaklines=false,
  literate=*{0}{{\color{blue}0}}1 {1}{{\color{blue}1}}1
             {2}{{\color{blue}2}}1 {3}{{\color{blue}3}}1
             {4}{{\color{blue}4}}1 {5}{{\color{blue}5}}1
             {6}{{\color{blue}6}}1 {7}{{\color{blue}7}}1
             {8}{{\color{blue}8}}1 {9}{{\color{blue}9}}1
}

\newenvironment{icompact}{
  \begin{list}{$\bullet$}{
    \parsep 0pt plus 1pt            % Controls the space between paragraphs within items
    \partopsep 0pt plus 1pt         % Controls the space between the first/last item and the surrounding text
    \topsep 2pt plus 2pt minus 1pt  % Controls the space before the first item and after the last item
    \itemsep 4pt plus 1pt           % Controls the space between items
    \parskip 2pt plus 1pt           % Controls the space between paragraphs within an item
    \leftmargin 0.13in              % Sets the left margin of the list
    \labelwidth 0.13in
    }}
  {\normalsize\end{list}}

\usepackage{caption}
\newcommand{\paragraphtitle}[1]{\vspace{5pt}\noindent\textbf{#1.}}

\begin{document}

\newcommand{\tool}{\textsc{Revelio}\xspace}
\def\link{\url{https://github.com/m1-llie/Revelio}}

\title{\tool: Cost-Efficient Agentic Memory Safety Vulnerability Detection \\ For Repository-Scale Codebases}

\author{
\IEEEauthorblockN{
Yiwei Hou,
Hao Wang,
Muxi Lyu,
Marius Momeu,\\
Eric Nguyen,
Taige Yang,
Koushik Sen,
Dawn Song,
David Wagner
}
\IEEEauthorblockA{
University of California, Berkeley\\
}
}

\maketitle

\IEEEpeerreviewmaketitle

\begin{abstract}
Memory safety vulnerabilities remain a significant threat even for projects with extensive fuzzing and manual auditing.
Recent results suggest that large language models hold great promise for detecting such vulnerabilities, but they are unreliable, at risk of hallucination, and challenging to scale to repository-size codebases.
This paper presents \tool, a cost-efficient end-to-end agentic framework for memory-safety vulnerability discovery.
\tool addresses the problem of hallucination by generating an executable Proof-of-Vulnerability, which is checked with a deterministic sanitizer.
It reduces cost using inexpensive LLMs and lightweight static analysis to help generate and rank vulnerability hypotheses, reporting vulnerabilities only when they can be reproduced and confirmed by a sanitizer.
We evaluated \tool on seven production-quality projects that had been continuously fuzzed for five to eight years, as well as on 100 randomly selected Arvo projects from the CyberGym benchmark.
With around one hour per project and a total cost of \$300, \tool discovered 19 previously unknown memory-safety vulnerabilities.
On benchmarks, \tool outperformed frontier coding agents across diverse backbone models at comparable token costs.
Our results suggest that \tool enables scalable and trustworthy end-to-end LLM-based memory-safety vulnerability detection.

\end{abstract}

\section{Introduction}

\begin{figure*}[t]
    \centering
    \includegraphics[width=\textwidth]{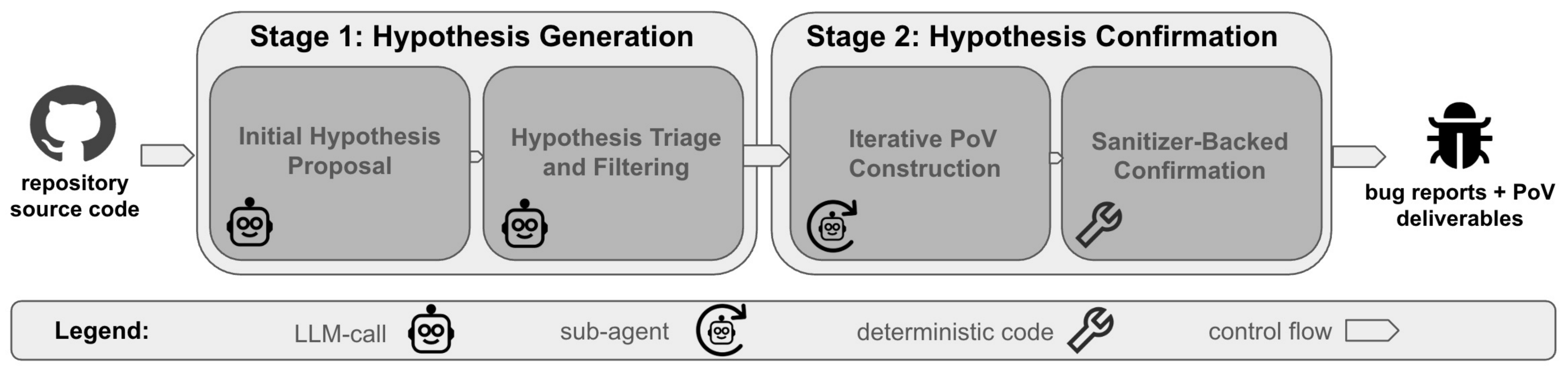}
    \caption{Overall workflow of \tool{} for end-to-end agentic memory safety vulnerability detection. An entry-point CLI gets the repository-scale codebase as input, and there's orchestration logic controlling the sub-stages.}
    \label{fig:workflow-overview}
\end{figure*}

\begin{figure}[t]
    \centering
    \includegraphics[width=\linewidth]{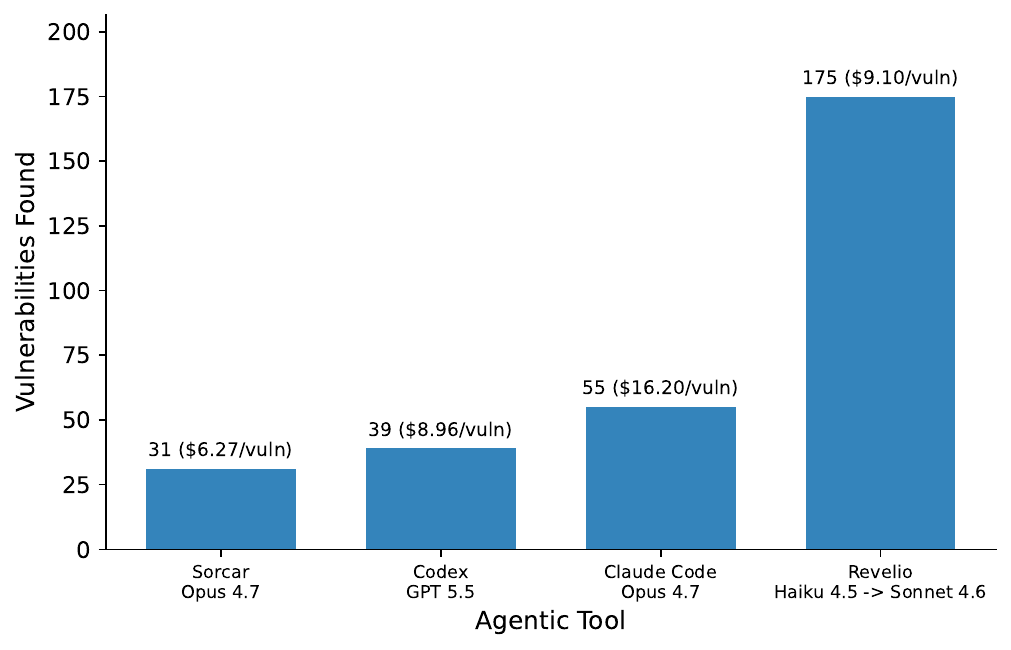}
    \caption{\tool{} is significantly more effective at detecting memory safety vulnerabilities than standard coding agents with a similar cost: \tool{} with Haiku 4.5 / Sonnet 4.6 outperforms Claude Code with Opus 4.7, Codex with GPT 5.5, and Sorcar with Opus 4.7.}
    \label{fig:tool_vulnerabilities_found}
\end{figure}

\noindent Memory safety vulnerabilities remain a serious threat to critical software systems.
Operating systems, browsers, network services, media libraries, and document parsers still rely heavily on C and C++, where subtle mistakes in bounds checks, lifetime management, integer arithmetic, and parser state can lead to exploitable memory errors.
Existing solutions such as fuzzing, static analysis, and manual security review have found many such bugs, yet vulnerabilities continue to be found in mature projects despite years of testing.
This makes memory safety vulnerability detection both practically important and technically difficult: easy bugs are often already removed, while remaining bugs require reasoning about rare input conditions, project-specific APIs, and hard-to-reach execution paths.

Recent advances in large language models (LLMs) and AI coding agents open new opportunities for vulnerability detection.
Industry systems and research prototypes increasingly show that agents can inspect repositories, reason about source code, and uncover real security vulnerabilities~\cite{openai2025aardvark,anthropic2026claudesecurity,liu2026agentflow}.
Compared with traditional static analysis, LLMs can interpret code semantics and generate hypotheses that are not tied to a fixed syntactic rule.
Compared with fuzzing alone, LLMs can reason about input formats, branch conditions, and API interactions before attempting to construct a triggering input.
These capabilities suggest that LLM-based systems may provide a useful new layer for discovering memory safety bugs in large codebases.

However, directly applying LLMs to repository-scale vulnerability detection remains challenging and expensive.
Earlier studies found LLMs performed poorly~\cite{ullah2024llms,yildiz2025benchmarkingllmagents,ding2025primevul,lin2025mammoth}.
The past year has seen a dramatic shift, with Anthropic's Project Glasswing demonstrating extraordinary progress in vulnerability detection, thanks to their latest model~\cite{anthropic2026mythosPreview,anthropic2026glasswing,anthropic2026glasswingUpdate}.
Recent academic efforts also report significant advances using sophisticated agent harnesses~\cite{guo2025repoaudit,zhao2026anypoc,he2026Co-RedTeam,liu2026agentflow}.
Unfortunately, existing methods suffer either from a substantial false positive rate (e.g., due to hallucinations or analysis errors)~\cite{guo2025repoaudit,zhao2026anypoc,steenhoek2025closing}, false negative rate (e.g., because models are not powerful enough to detect some vulnerabilities, or they don't have sufficient coverage of vulnerability types)~\cite{ullah2024llms,ding2025primevul,yildiz2025benchmarkingllmagents,li2024iris,lekssays2025llmxcpg}, or high cost (e.g., because the latest advanced models are very expensive to run on an entire repository)~\cite{guo2025repoaudit,yildiz2025benchmarkingllmagents,yang2025knighter,widyasari2026vultrial}. 

In this paper, we develop a system for detecting memory safety vulnerabilities.
We show that it is able to detect new vulnerabilities in well-vetted software projects, at a relatively low cost.
The system is designed to avoid false positives, avoiding burdening humans with hallucinations and false reports.
Our experiments with known vulnerabilities in benchmarks suggest that it can find most memory safety vulnerabilities.
This indicates that LLM-powered vulnerability detection is very effective, even without using non-public models like Mythos Preview~\cite{anthropic2026mythosPreview} and without complex harness orchestrations.

\paragraphtitle{Insight}
Our key insight is that memory safety vulnerability detection can be decomposed into two qualitatively different tasks.
The first task is code review: scan many files and identify possible memory safety bugs (hypotheses).
Code review can tolerate errors, as long as it errs on the side of producing too many hypotheses (wrongly accusing code of being buggy, even if it is actually safe), because false hypotheses can be filtered later.
The second task is executable confirmation: converting a promising hypothesis about a possible vulnerability into an input that demonstrates the existence of the vulnerability.
This second task requires stronger reasoning and iterative debugging, but it is applied only to a much smaller post-triage candidate set.
Moreover, its results can be deterministically verified using existing sanitizers.
Therefore, we can use inexpensive models for broad code review to reduce false negatives, and stronger models and deterministic verification for confirmation to eliminate false positives, significantly reducing the cost of vulnerability detection.

\paragraphtitle{Approach}
Based on this insight, we design \tool, an end-to-end agentic framework for detecting memory safety vulnerabilities in repository-scale C/C++ codebases.
As shown in Figure~\ref{fig:workflow-overview}, \tool{} follows a simple two-stage design.
In the Hypothesis Generation stage, \tool{} scans source files, performs lightweight static preprocessing, and uses cheap LLM models to generate hypotheses about candidate memory safety vulnerabilities.
This stage is designed for coverage: it searches for possible bugs, but some hypotheses might be false alarms.
In the Hypothesis Confirmation stage, \tool{} attempts to turn the highest-ranked hypotheses into executable evidence.
For each candidate, \tool{} selects an executable test harness (program entry point) that can reach the relevant code, invokes a stronger model to synthesize an input that demonstrates the vulnerability, also known as the ``Proof-of-Vulnerability''(PoV).
We then validate the PoV by executing the program with a sanitizer enabled.
\tool{} reports a vulnerability only when this PoV reproducibly triggers a detection from the sanitizer.
This design establishes a strict evidence boundary: LLMs propose and construct, while program execution and the sanitizer serve as a trustworthy verifier.

\paragraphtitle{Results}
We first apply \tool{} to randomly selected production-quality projects that have been continuously fuzzed by OSS-Fuzz~\cite{serebryany2017ossfuzz} for five to eight years.
\tool{} discovers 19 previously unknown memory safety vulnerabilities, including seven CVEs, with a median runtime of about one hour per project at a cost of about \$42 per project.

We also evaluate \tool{} on 100 randomly selected Arvo projects~\cite{mei2024arvo} from CyberGym~\cite{wang2025cybergym}, and compare it to several general-purpose AI coding agents.
As shown in Figure~\ref{fig:tool_vulnerabilities_found}, \tool{} finds substantially more vulnerabilities than Claude Code, Codex, and Sorcar (175 vs.\ 55, 39, and 31).
Moreover, \tool has zero false positives, while the other coding agents have a substantial false positive rate.
\tool{} is also significantly more effective at detecting the known vulnerabilities in these CyberGym projects.
These experiments show that our two-stage workflow is very effective and cost efficient at detecting memory safety vulnerabilities, and that we don't need access to Mythos to detect lots of vulnerabilities (existing models suffice).

\paragraphtitle{Contribution} In summary, our contributions are threefold:

\begin{icompact}
    \item We show that memory safety vulnerabilities can be detected at scale with a two-stage process: code review with a cheaper model, followed by confirmation with a stronger model.
    We propose an agentic system design that scales up to large codebases with high vulnerability detection capability, high detection recall, zero false positives, and acceptable cost.

    \item We design and implement \tool, an end-to-end agentic vulnerability detection framework intended for practical deployment. \tool is available at \link.

    \item We evaluate \tool{} on well-tested real-world software projects, discovering 19 previously unknown vulnerabilities at approximately \$300 total cost. 
    It also outperforms advanced AI coding agent baselines on benchmark evaluations.
\end{icompact}

\section{Background}

\vspace{-5pt}
\paragraphtitle{Memory Safety Vulnerabilities and Sanitizers}
Memory safety vulnerabilities remain a major threat to C/C++ software and systems, spanning various CWE classes: spatial violations such as CWE-125 Out-of-Bounds Read~\cite{cwe125} and CWE-787 Out-of-Bounds Write~\cite{cwe787}; temporal violations such as CWE-416 Use-After-Free~\cite{cwe416} and CWE-415 Double Free~\cite{cwe415}; and uninitialized-memory use, such as CWE-457 Use of Uninitialized Variable~\cite{cwe457}.
Modern compilation pipelines use runtime instrumentation tools such as AddressSanitizer (ASan)~\cite{serebryany2012addresssanitizer}, MemorySanitizer (MSan)~\cite{stepanov2015memorysanitizer}, and UndefinedBehaviorSanitizer (UBSan)~\cite{llvm2013ubsan} to detect memory errors and undefined behavior during execution. In automated vulnerability discovery, sanitizer reports provide concrete, reproducible evidence of invalid runtime behavior, reducing false positives from speculative analysis~\cite{serebryany2017ossfuzz}.

\paragraphtitle{Software Vulnerability Detection}
Existing vulnerability detection has largely relied on static program analysis and dynamic testing. Static analysis identifies suspicious memory-management patterns at scale~\cite{zhang2025crossentry}, while fuzzing and sanitizer-guided testing provide concrete evidence of reachable bugs~\cite{serebryany2017ossfuzz}.
LLM-based agents offer a promising approach to repository-scale vulnerability discovery by combining code reasoning with iterative tool use~\cite{guo2025repoaudit}. However, achieving broad coverage and turning model-generated hypotheses into reproducible evidence remain resource-intensive challenges~\cite{wang2025cybergym,zhao2026anypoc}. \tool addresses these challenges through a cost-effective workflow that first prioritizes promising hypotheses and then confirms them with sanitizer-validated PoVs.

\section{Design of \tool{}}
\label{sec:methodology}

\noindent  \tool is designed around several design goals:
\begin{icompact}
\item \textbf{Thorough coverage of the codebase.}
Our goal is to find all memory safety vulnerabilities in a codebase, or as many as possible.
Fuzzing typically can only find some vulnerabilities, and thorough manual security review is expensive.
We mimic a thorough manual review, using LLMs to scan every file of the codebase.
We use an inexpensive model to identify all potentially vulnerable parts of the code, and then apply more detailed reasoning to validate each potential vulnerability.

\item\textbf{Avoid false alarms.}
LLMs are notorious for hallucination, and AI vulnerability detection can also hallucinate fake vulnerability reports.
To avoid bothering maintainers with erroneous claims, our goal is that every discovery must be verified with a deterministic verifier.
If we can find an input that causes a sanitizer to report a vulnerability when that input is executed (a PoV), we can have high confidence in the discovery.
It also produces reproducible evidence of the vulnerability for developers, which hopefully makes the vulnerability report easier to trust and to fix.

\item\textbf{Cost efficiency.} We use different model tiers for different tasks.
Cheaper models are sufficient for scanning code to generate hypotheses about potential vulnerabilities, as long as we don't mind some false alarms.
More capable models are applied for expensive reasoning tasks such as PoV construction and iterative debugging.
This division of labor allows \tool{} to scale to large codebases while concentrating expensive inference only on promising candidates.
\end{icompact}

\subsection{Overall Workflow}
\noindent The design of \tool{} is first motivated by a simple observation:
\textit{large language models are increasingly effective at reasoning about source code, but their natural-language vulnerability claims are not reliable enough.}
A practical vulnerability detection system must therefore separate speculative reasoning from executable confirmation.
As shown in Figure~\ref{fig:workflow-overview}, \tool{} follows this principle by decomposing detection into two stages: Hypothesis Generation and Hypothesis Confirmation.

\begin{figure*}[t]
    \centering
    \includegraphics[width=0.9\textwidth]{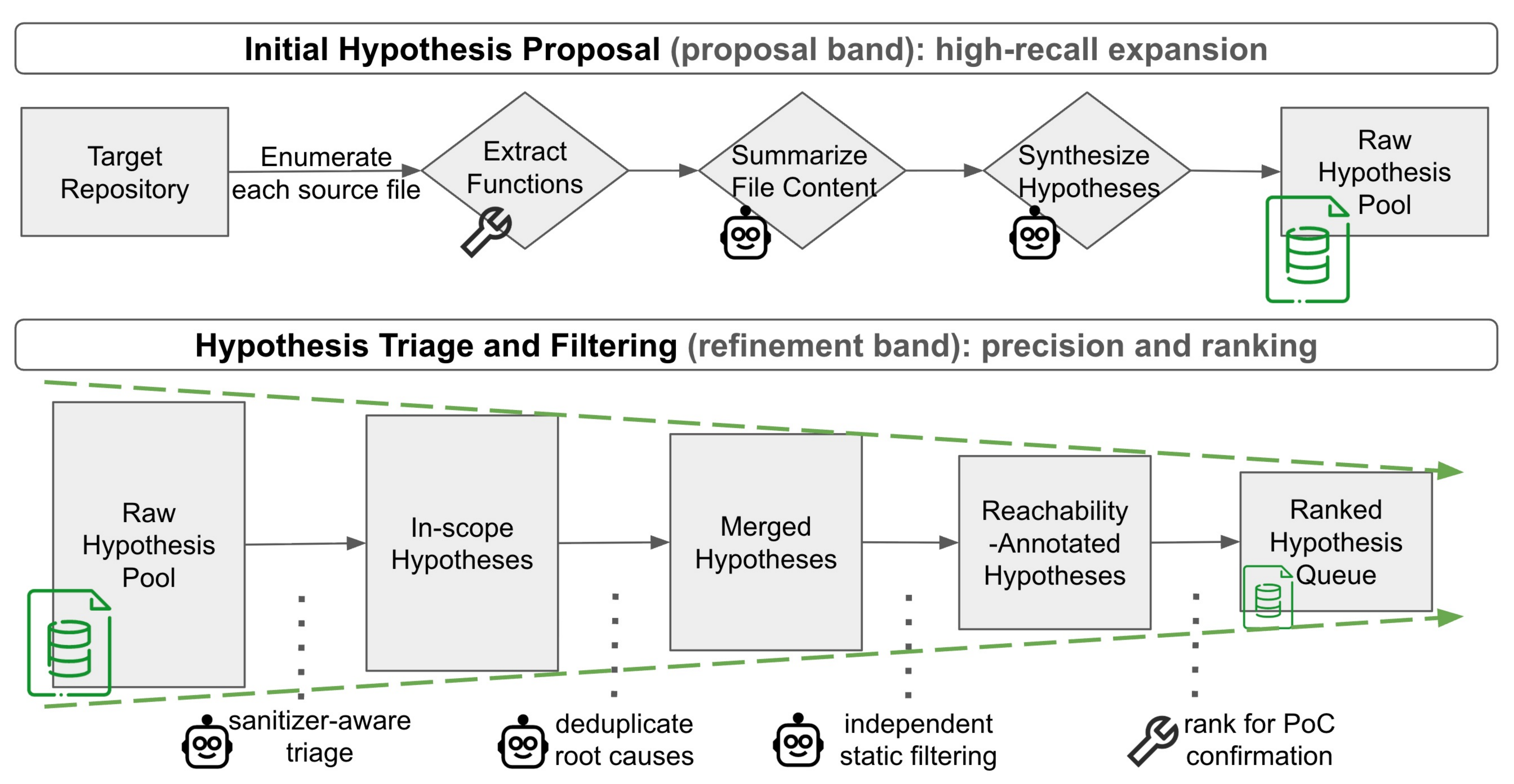}
    \caption{Hypothesis Generation Stage pipeline. \tool{} first expands the target repository into a high-recall pool of raw memory safety vulnerability hypotheses, then refines this pool through a funnel-style filtering before handing structured candidates to the Hypothesis Confirmation stage. Boxes denote intermediate artifacts, while diamonds denote the operations that transform one artifact into the next.}
    \label{fig:hypothesis-generation-stage}
\end{figure*}

\subsubsection{Hypothesis Generation} 
In the first stage, \tool{} scans each file with a sub-agent to generate hypotheses about candidate memory safety vulnerability.
Each hypothesis identifies a potentially vulnerable code region, describes the suspected root cause, and explains the conditions under which an attacker-controlled input may trigger memory-unsafe behavior.
Triage and filtering then filter the list to those with higher confidence.
The first stage is designed for high coverage and high recall: we intentionally prioritize the number of diverse candidate hypotheses even though many of them will ultimately be proven wrong.
This is because missing a subtle vulnerability during the initial scan is more costly than filtering an incorrect hypothesis later.
They are not all reliable findings, but the second stage gives us a way to tell which ones are real, and the exhaustive search helps us avoid missing vulnerabilities.
In summary, inspired by The Bitter Lesson~\cite{sutton2019bitter}, we lean into scale:
\textit{rather than limiting the search space with handcrafted heuristics, we take advantage of cheap models to examine everything.}

\subsubsection{Hypothesis Confirmation}
In the second stage, \tool{} attempts to verify each surviving hypothesis by creating a PoV, i.e., an input that triggers a sanitizer error.
For memory safety vulnerabilities, code sanitizers provide reliable validation. 
A candidate is considered confirmed if and only if the generated PoV triggers a sanitizer-detected failure in an actual target binary.
In this stage, \tool{} performs iterative trial-and-error PoV generation to generate an input that triggers the vulnerability, and then verifies that running the program on this input triggers a code sanitizer error report.
Verification is deterministic and reliable, and any test case that triggers a code sanitizer error is typically a high-risk issue, so with this verification process, we can be confident that \tool{} is finding real issues.
Additionally, the generated PoV input enters the software through existing codebase-provided test harnesses that encode valid preconditions and exercise publicly accessible attack surfaces.

\subsection{Vulnerability Hypothesis Generation}
\noindent The first stage converts a target codebase into a ranked set of vulnerability hypotheses.
This stage is designed for breadth and codebase coverage:
rather than asking the agentic system to directly decide whether a project contains a vulnerability, \tool{} analyzes each file to generate concrete, localized, and testable claims about possible memory safety vulnerabilities.
The stage is organized as a funnel-style pipeline, as shown in Figure~\ref{fig:hypothesis-generation-stage}.
\tool{} first produces a broad pool of raw hypotheses, then removes candidates that are out of scope, duplicated, or statically invalid, and finally ranks the remaining candidates.
This design minimizes the risk of missing real vulnerabilities, while reducing the number of vulnerability claims that need to be confirmed by the second stage.

\subsubsection{Initial Hypothesis Proposal}
Given a target project, \tool{} first enumerates C/C++ source files inside the codebase.
It excludes directories that belong to testing infrastructure (e.g., fuzzing or unit test related).
For each selected file, \tool{} reads the source and prepares a per-file analysis context.

Before invoking the LLM model, \tool{} performs a lightweight static preprocessing.
It parses the code, extracts function boundaries from the AST, and batches functions by size so that the model can analyze both whole-file behavior and individual function bodies.
It also runs an intraprocedural static analysis to identify whether function parameters are guarded by NULL checks, bounds checks, assertions, validation calls, or related conditions.
This analysis does not decide whether a vulnerability exists; instead, it gives compact evidence about which parameters appear unchecked and therefore deserve closer scrutiny.

\tool{} then performs multiple LLM analysis passes over each file.
The first pass asks a LLM to summarize the file's functionality and attribute code regions to high-level features.
This summary is retained as context for later analysis, but it does not produce vulnerability hypotheses.
Subsequent passes generate hypotheses from different perspectives: whole-file feature interactions, possible memory safety conditions, unchecked argument patterns, and per-function local analysis.
These hypothesis-generation passes emit candidate hypotheses as JSON records in a structured schema containing a short textual vulnerability summary, a list of code locations (file, function, consecutive range of line numbers) where it occurs, potential preconditions, and notes about the type of vulnerability.

This multi-pass design intentionally favors recall.
A subtle memory safety vulnerability may be visible only under a particular framing:
for example, as a parser-state inconsistency, an unchecked size field, or an incorrect branch condition.
Running several focused passes gives the model multiple opportunities to notice the same bug mechanism from different angles.
However, this also produces duplicates and low-confidence candidates, so \tool{} treats these raw LLM model outputs only as vulnerability hypotheses, not trustworthy final findings.

\subsubsection{Hypothesis Triage and Filtering}
\begin{table}[!t]
\centering
\caption{Hypothesis artifact schema produced by the Hypothesis Generation Stage.}
\label{tab:hypothesis-artifact-schema}
\footnotesize
\setlength{\tabcolsep}{4pt}
\renewcommand{\arraystretch}{1.15}
\begin{tabularx}{\columnwidth}{
>{\raggedright\arraybackslash\hsize=0.72\hsize}X
>{\raggedright\arraybackslash\hsize=1.25\hsize}X
>{\raggedright\arraybackslash\hsize=1.03\hsize}X}
\toprule
\textbf{Field} & \textbf{Description} & \textbf{Example} \\
\midrule
\texttt{hypo\_id} & Unique identifier assigned after filtering and ranking & \texttt{H03} \\
\texttt{title} & Short vulnerability claim & ``Heap out-of-bounds read in image parser'' \\
\texttt{description} & Explanation of the suspected root cause & ``A crafted chunk length may exceed the decoded buffer size before bounds validation.'' \\
\texttt{file\_path} & Source file containing the vulnerability & \texttt{src/parser.c} \\
\texttt{function} & Primary function associated with the hypothesis & \texttt{parse\_chunk} \\
\texttt{references} & Code locations that support the hypothesis & \texttt{src/foo.c:21--28} \\
\texttt{severity} & Estimated impact used for prioritization & \texttt{high} \\
\texttt{primitive} & Type of vulnerability & \texttt{oob-read}, \texttt{oob-write}, \texttt{uaf}, \texttt{int-overflow} \\
\shortstack[l]{\texttt{attacker\_}\\\texttt{controls}} & Whether attacker-controlled input reaches the bug & \texttt{input} \\
\texttt{sanitizers} & Expected sanitizer oracle & \texttt{asan}, \texttt{ubsan}, \texttt{msan} \\
\texttt{cwe\_ids} & Normalized weakness category & \texttt{CWE-125} \\
\texttt{reachable} & Whether a test harness links the relevant function & \texttt{true}, \texttt{false}, \texttt{unknown} \\
\texttt{harnesses} & Candidate test harnesses & \texttt{["parser\_driver"]} \\
\texttt{confidence} & Sub-agent's confidence & \texttt{0.82} \\
\bottomrule
\end{tabularx}
\end{table}

After the initial proposal step, \tool{} converts the noisy raw hypothesis pool into a smaller set of structured candidates for PoV construction.
It first triages these hypotheses by determining whether each hypothesis describes an in-scope memory safety issue, whether attacker-controlled input reaches the alleged bug, which type of vulnerability it implies, which sanitizer should detect it, and which CWE category is most relevant.
For example, pure code-quality issues, failures of graceful error handling, caller-contract violations, and non-memory-safety logic bugs are filtered out.
We focus specifically on memory safety vulnerabilities that can be triggered by outside input execution via an existing executable test harness.

\tool{} then deduplicates overlapping hypotheses.
It first identifies candidate duplicate pairs using line-range overlap and shared CWE labels, then uses a LLM to determine whether each pair describes the same underlying root cause.
When duplicates are merged, we keep the hypothesis that has more context about relevant locations in the code.
Deduplication avoids wasting effort confirming multiple natural-language descriptions of the same vulnerability.

Next, \tool{} annotates each hypothesis to identify which test harnesses can plausibly reach the relevant function. 
The software we analyze comes with one or more test harnesses; each test harness is an entry point into the code, and specifically, a runnable program that invokes one or more functions in the code on some input.
Therefore, if a test harness transitively reaches (invokes) the potentially vulnerable code, we can confirm the existence of the vulnerability by searching for an input to the test harness that triggers the vulnerability.
For each top-ranked hypothesis, \tool{} identifies candidate test harnesses by finding corresponding binaries whose symbol tables contain the hypothesis's relevant function. 

The final output of the Hypothesis Generation Stage is a ranked list of structured hypotheses, ordered by reachability, severity, and confidence.
Table~\ref{tab:hypothesis-artifact-schema} summarizes the information produced about each hypothesis.
% These fields are used both for ranking and for handoff to the next stage, Hypothesis Confirmation. 
In particular, the \textit{function}, \textit{references}, \textit{sanitizers}, and \textit{harnesses} fields help the next stage identify code segments that are relevant, the expected failure mode, and which test harness to use to confirm the vulnerability.

\subsection{Hypothesis Confirmation by PoV Construction}
\noindent The second stage determines whether a hypothesis corresponds to a real, exploitable memory safety vulnerability.
This stage deliberately has a narrower focus than hypothesis generation: \tool{} reports a vulnerability only if it can generate an input that triggers a sanitizer-detected failure in the target program through a realistic and reachable attack surface from the outside.
While the Hypothesis Generation stage searches broadly for plausible memory safety vulnerability hypotheses, the Hypothesis Confirmation stage tests each hypothesis via some test harness to see if it can be validated by a code sanitizer. 

\subsubsection{Iterative PoV Construction}
\begin{table}[!t]
\centering
\caption{How Stage~1 hypothesis fields are used by PoV construction in Stage~2.}
\label{tab:stage-handoff-pov-confirmation}
\footnotesize
\setlength{\tabcolsep}{4pt}
\renewcommand{\arraystretch}{1.15}
\begin{tabularx}{\columnwidth}{
>{\raggedright\arraybackslash\hsize=0.70\hsize}X
>{\raggedright\arraybackslash\hsize=0.95\hsize}X
>{\raggedright\arraybackslash\hsize=1.35\hsize}X}
\toprule
\textbf{Hypothesis Field} & \textbf{Used By PoV Stage For} & \textbf{Why It Matters} \\
\midrule
\texttt{description}, \texttt{primitive} & Trigger reasoning & Explains the suspected bug mechanism \\
\texttt{function} & Executable test harness discovery via symbol lookup & Selects harnesses that can reach the suspected code \\
\texttt{file\_path}, \texttt{references} & Source inspection & Directs the PoV agent to relevant code locations \\
\texttt{sanitizers} & Oracle selection & Indicates whether ASAN, UBSAN, or MSAN should detect the bug \\
\texttt{severity}, \texttt{confidence}, \texttt{reachable} & Ranking before PoV attempts & Allocates expensive PoV budget to promising candidates \\
\texttt{harnesses} & Test harness selection & Avoids asking the model to guess the correct executable \\
\bottomrule
\end{tabularx}
\end{table}

\tool{} invokes an agent instance to construct a PoV input that exercises the vulnerability.
The agent receives a structured context packet containing the selected hypothesis, affected function, code references, selected test harness, etc.
As shown in Table~\ref{tab:stage-handoff-pov-confirmation}, this context packet provides necessary information for the agent to construct a valid PoV input, while saving effort on the agent for LLM reasoning context window management.
\tool{} provides the agent with three tools: a \texttt{shell} tool for inspecting the repository and writing files, a \texttt{validation} tool for testing a candidate PoV, and a \texttt{finish} tool for submitting structured results.
The agent is instructed to inspect the executable harness, infer the expected input format, write a Python script that generates raw input bytes of the PoV, and iteratively refine the input based on previous validation's feedback.

A valid PoV must exercise a publicly accessible attack surface, as represented by the test harness.
This rejects vulnerabilities that can only be triggered with privileged local access, direct mutation of internal program state, calls to private/internal APIs, debugger intervention, source-code modification, or unrealistic preconditions that an external attacker could not satisfy.
This constraint is important because  a crash caused by violating an internal API contract is not necessarily an exploitable vulnerability in the target's real attack surface.

\subsubsection{Sanitizer-Backed Confirmation}
The \texttt{validation} tool available to the agent provides a trustworthy oracle for the Hypothesis Confirmation stage.
The tool runs the test harness on the input with one or more sanitizers enabled, e.g., AddressSanitizer, UndefinedBehaviorSanitizer, and/or MemorySanitizer.
The tool parses return codes and known crash signatures to determine whether a crash occurred.
If the input did not trigger a sanitizer crash, the tool returns sanitizer output to the agent, so it can iteratively refine the input.

This provides trustworthy, programmatic verification of vulnerability claims.
We do not blindly trust any AI agent claiming there is a vulnerability, as that can have false positives.
After the agent submits a final PoV, the system independently re-executes the generated input in a fresh subprocess and verifies the sanitizer report programmatically.
This prevents cases where an agent overstates success, misinterprets logs, reports a stale crash, or constructs an output that appears valid only inside its own trajectory.
The agent may try candidate inputs, reason about parser constraints, and iterate on failed attempts, but it cannot declare success by assertion.
A hypothesis is confirmed only when the sanitizer reports a violation during independent re-execution.
If all validation attempts fail, \tool{} moves on to the next vulnerability hypothesis.

Once a crash is confirmed, \tool{} invokes the final reporting component, the reporter sub-agent.
The reporter agent receives the original hypothesis, the generated PoV script, the raw PoV input path, the reproduction command, and sanitizer evidence.
It writes a developer-facing report containing the vulnerability details, affected file and function, triggering input, sanitizer output, and reproduction steps.
The report is emitted only after confirmation, so the final artifact set contains executable, trustworthy evidence.

\section{Implementation}
\noindent We implemented \tool{} in 4.1K lines of Python code.
The system architecture decouples high-level orchestration from agent logic and execution environments.

\paragraphtitle{Agentic System Architecture and Orchestration}
Each sub-agent of the system is instantiated from a \texttt{DefaultAgent} class, which encapsulates a specific LLM model, an execution environment, a prompt template, and a list of tools.
We define prompts and configurations as YAML-based Jinja templates.
At runtime, the orchestrator dynamically populates these templates with contextual variables, such as \textit{project\_path} (the filesystem root of the target under test), \textit{hypothesis\_id} (a unique identifier for the current vulnerability candidate), \textit{context\_packet} (aggregated metadata from previous analysis stages), and \textit{executable\_harness} (the specific entry point identified for testing).
All prompts used in this work are shown in Appendix~\ref{app:prompts}.

\paragraphtitle{Static Preprocessing in Hypothesis Generation}
We parse the codebase with the tree-sitter library and implement a custom, lightweight C/C++ analyzer to generate parameter validation summaries.
These analyses are intentionally shallow; rather than attempting to formally prove safety or vulnerability, which is computationally expensive, they provide a few compact hints to the agent.
This allows the LLM to efficiently filter the hypothesis space without being overwhelmed by verbose static analysis logs.

\paragraphtitle{Execution and Validation Environment}
We execute each codebase-under-test in a Docker container to ensure environment isolation and reproducibility.
Each PoV is tested by running the code within this container.

\section{Evaluation}
\label{sec:evaluation}

\vspace{-5pt}
\paragraphtitle{Research Questions}
We evaluate \tool{} through the following research questions:

\begin{icompact}

    \item \textbf{RQ1 (real-world vulnerability discovery, Section~\ref{sec:real-world-0day}).}
    Can \tool{} discover previously unknown vulnerabilities in real-world, production-quality software at maintainer-deployable cost?
    
    \item \textbf{RQ2 (detection effectiveness and cost efficiency, Section~\ref{sec:baselines-comparison}).}
    How effective and cost-efficient is \tool{} for end-to-end detection of known memory safety vulnerabilities, compared with advanced general-purpose AI coding agents?

    \item \textbf{RQ3 (ablations, Section~\ref{sec:harness-vs-prompting}).}
    How effective are the individual components of \tool? How much of the performance gain can be obtained through prompting alone, and how much requires programmatic harness enforcement?
\end{icompact}

\paragraphtitle{Evaluation design}
We begin with a real-world zero-day hunt to test whether \tool{} produces externally meaningful findings in production-level software.
We then move to controlled evaluation on a dataset of known vulnerabilities, so we can measure how many vulnerabilities are missed and support direct comparison against general-purpose AI coding agents on effectiveness and cost efficiency.
We finally evaluate the impact of different aspects of our design, to identify the benefit of each component.

\paragraphtitle{Configurations}
We use these model pinned snapshots in agent comparison experiments:
Claude Haiku 4.5 (\texttt{claude-haiku-4-5-20251001}),
Claude Sonnet 4.6 (\texttt{claude-sonnet-4-6}),
Claude Opus 4.7 (\texttt{claude-opus-4-7});
GPT 5.4 mini (\texttt{gpt-5.4-mini}), GPT 5.4 (\texttt{gpt-5.4}), and GPT 5.5 (\texttt{gpt-5.5-2026-04-23}).
We choose these models to represent three LLM capability and cost tiers: small/low-cost, mid-tier, and frontier models.
\tool's default configuration in experiments uses Claude Haiku 4.5 for Hypothesis Generation and Claude Sonnet 4.6 for Hypothesis Confirmation.

\paragraphtitle{Cost accounting}
We report total token usage and billed API cost in dollars over all model calls.
Appendix~\ref{app:cost} lists the rates, cached-token treatment, and API-cost calculation.
Prompt caching is enabled for all systems whose API supports it, and cached reads and writes are priced using the provider-specific billing categories.
Per-experiment costs are cross-validated against provider billing dashboards before reporting.

\paragraphtitle{Machine}
All experiments were conducted on an Ubuntu machine equipped with an Intel(R) Xeon(R) Gold 5320 CPU @ 2.20GHz processor, 503GB of memory, and 77T of storage.
The LLM model requests were accessed through the public network.

\subsection{Real-World Zero-day Vulnerability Detection} \label{sec:real-world-0day}
\noindent We first evaluate whether \tool{} can discover real previously unknown vulnerabilities, i.e., zero-days, in production-quality codebases end-to-end.
Results are briefly summarized in Table~\ref{tab:0day-brief}.

\begin{table}[!t]
\centering
\caption{Brief introduction of zero-day vulnerability detection performance.}
\label{tab:0day-brief}
\small
\begin{tabularx}{\columnwidth}{@{}>{\raggedright\arraybackslash}X r@{}}
\toprule
\textbf{Metric}  & \textbf{Count} \\
\midrule
Targeted Oss-Fuzz project codebases    & 7 \\
Zero-day vulnerabilities found  & 19 \\
CVE ID assigned / requested & 7 \\
Median tokens (In/Out)  & 14.6M / 1.3M \\
Median token cost  & \$42 \\
Median time duration for \tool running    & 65min \\
Total cost for zero-day hunting evaluation &  \$300 \\
\bottomrule
\end{tabularx}
\end{table}

\begin{table}[!t]
\centering
\caption{Project information for the OSS-Fuzz zero-day hunt.}
\label{tab:e4-projects}
\footnotesize
\setlength{\tabcolsep}{4pt}
\begin{tabularx}{\columnwidth}{@{}l >{\raggedright\arraybackslash}X r r@{}}
\toprule
Project        & Domain                & LoC (k) & OSS-Fuzz   \\
               &                       &         & tenure (y) \\
\midrule
DNSMasq        & DNS / DHCP / TFTP     & 25  & 5 \\
OpenEXR        & HDR image library     & 58  & 6 \\
assimp         & 3D model import       & 125 & 6 \\
cairo          & 2D vector graphics    & 176 & 6 \\
The Sleuth Kit & Digital forensics     & 218 & 6 \\
libheif        & HEIF/AVIF codec       & 559 & 7 \\
Poppler        & PDF rendering         & 911 & 8 \\
\bottomrule
\end{tabularx}
\end{table}

We randomly select seven projects from OSS-Fuzz~\cite{serebryany2017ossfuzz} for the zero-day hunt (Table~\ref{tab:e4-projects}).
Projects under OSS-Fuzz usually have been continuously fuzzing tested for years.
The seven selected projects span a variety of domains, including network services, image and document processing, 2D/3D graphics, media decoding, and digital forensics.
These projects contain various input-processing workloads, including protocol handling, structured file parsing, graphics rendering, and codec implementation, all of which are common sources of memory safety errors.
The selected projects vary in size from 25k to 911k LoC and have been integrated into OSS-Fuzz for between five and eight years.
Therefore, they provide a challenging and diverse set of targets for evaluating whether \tool{} can generalize to heterogeneous, mature, and extensively fuzzed real-world software projects.

\begin{table*}[!t]
\centering
\caption{Results of the \tool{} zero-day hunt across seven mature OSS-Fuzz projects that had already undergone five to eight years of continuous fuzzing. \tool{} discovered 19 previously unknown memory-safety vulnerabilities, all of which were manually validated and responsibly disclosed; the table summarizes each finding together with its estimated severity and the corresponding maintainer response.}
\label{tab:vuln-info}
\scriptsize
\setlength{\tabcolsep}{3pt}
\begin{tabularx}{\textwidth}{l l l X l l l l}
\toprule
ID & Project & Version & Finding & Vulnerability Class & CVSS v3.1 Severity & Report Time & Maintainer Response \\
\midrule
1 & DNSMasq & v2.93test9 & DS record log buffer overflow & Heap-based buffer overflow & High & April 18, 2026 & Fixed, CVE requested \\
\midrule
2 & \multirow{2}{*}{OpenEXR} & \multirow{2}{*}{v3.4.10} & ID manifest shift exponent overflow & Integer overflow & Critical & April 20, 2026 & Fixed, CVE-2026-42217 \\
3 & & & Manifest string prefix OOB read & Out-of-bounds read & Critical & April 20, 2026 & Fixed, CVE-2026-42216 \\
\midrule
4 & \multirow{2}{*}{Poppler} & \multirow{2}{*}{v26.04.0} & Image stream width multiplication overflow & Integer overflow & Moderate & April 21, 2026 & Fixed, CVE requested \\
5 & & & TIFF predictor bit-shift exponent UB & Integer overflow & Moderate & April 21, 2026 & Fixed, CVE requested \\
\midrule
6 & assimp & v6.0.4 & Empty URI zero-advance abort & Integer underflow & Moderate & April 7, 2026 & Pending \\
\midrule
7  & \multirow{5}{*}{cairo} & \multirow{5}{*}{v1.18.4} & Fixed-point floor negation overflow & Integer overflow & Moderate & April 21, 2026 & Pending \\
8  & & & Path slope subtraction integer overflow & Integer overflow & Moderate & April 21, 2026 & Pending \\
9  & & & Stroke extent addition integer overflow & Integer overflow & Moderate & April 21, 2026 & Pending \\
10 & & & Embedded font name OOB read & Heap-based buffer overflow & High & April 21, 2026 & Pending \\
11 & & & Type1 font metric multiplication overflow & Integer overflow & Moderate & April 21, 2026 & Pending \\
\midrule
12 & \multirow{5}{*}{libheif} & \multirow{5}{*}{v1.21.2} & Track chunk count OOB access & Heap-based buffer overflow & Moderate & April 20, 2026 & Fixed, CVE-2026-47254 \\
13 & & & Component ID wraparound OOB write & Out-of-bounds write & Moderate & April 20, 2026 & Fixed \\
14 & & & Aux sample-chunk mismatch OOB & Out-of-bounds read & High & April 20, 2026 & Fixed, CVE-2026-41071 \\
15 & & & Tile count wraparound OOB crash & Heap-based buffer overflow & Moderate & April 20, 2026 & Fixed \\
16 & & & Track array unchecked size overflow & Heap-based buffer overflow & Moderate & April 20, 2026 & Fixed \\
\midrule
17 & \multirow{3}{*}{The Sleuth Kit} & \multirow{3}{*}{v4.15.0} & Signed offset bypass OOB read & Heap-based buffer overflow & High & April 18, 2026 & Pending \\
18 & & & Inodes-per-block OOB 128-byte read & Heap-based buffer overflow & High & April 18, 2026 & Pending \\
19 & & & Index update-sequence array OOB & Out-of-bounds read & Moderate & April 18, 2026 & Pending \\
\bottomrule
\end{tabularx}
\end{table*}

\tool{} found 19 vulnerabilities in these projects (see Table~\ref{tab:vuln-info}), all of which were manually validated and then submitted to project maintainers for responsible disclosure.
\tool{} was configured to propose five hypotheses for each source file in the Hypothesis Generation stage and try five times to construct a PoV for each vulnerability hypothesis.

\begin{table}[!t]
\centering
\caption{Vulnerability class distribution of \tool's 19 confirmed findings, categorized by CWE of the root cause.}
\label{tab:e4-classes}
\small
\begin{tabular}{l r}
\toprule
Vulnerability class                                        & Count \\
\midrule
CWE-190: Integer Overflow or Wraparound                    & 7 \\
CWE-122: Heap-based Buffer Overflow          & 7 \\
CWE-125: Out-of-bounds Read                 & 3 \\
CWE-191: Integer underflow (Wrap or Wraparound)     & 1 \\
CWE-787: Out-of-bounds Write              & 1 \\
\midrule
\textbf{Total}                                   & \textbf{19} \\
\bottomrule
\end{tabular}
\end{table}

\paragraphtitle{Vulnerability Types and Root Causes}
As shown in Table~\ref{tab:e4-classes}, \tool{} can find many different types of memory safety vulnerabilities.
Arithmetic-related defects dominate the findings: integer overflows and integer underflow together account for a large fraction of the confirmed vulnerabilities.
This suggests that many memory safety violations are not caused by direct misuse of memory primitives alone, but are triggered earlier by incorrect size, offset, or length computations, which subsequently lead to unsafe allocations or boundary checks.
Buffer-bound violations are also prevalent, including both heap-based buffer overflows and out-of-bounds accesses.
These vulnerabilities typically arise when attacker-controlled or malformed inputs influence buffer sizes, indices, or copy lengths, causing the program to read or write beyond the intended memory region.
In particular, the presence of both out-of-bounds reads and writes indicates that \tool{} is able to uncover vulnerabilities with different security impacts, ranging from information disclosure to memory corruption and potential control-flow hijacking.
Overall, the distribution shows that \tool{} is effective not only at detecting explicit memory access errors, but also at reasoning about the arithmetic and data-flow conditions that lead to memory-unsafe behavior.

\paragraphtitle{Vulnerability Severity}
After manual validation and confirmation, we believe that all the discovered vulnerabilities have the potential to cause security related consequences.
Regarding the vulnerabilities that have been confirmed and fixed by project maintainers, seven have been assigned or requested as CVEs.
We manually evaluated all 19 vulnerabilities, and we estimate that about half are likely to be low severity, e.g., DoS, and the remainder are more serious.
We also evaluated all 19 using CVSS 3.1; CVSS classifies two as critical, five as high, and 12 as moderate.

\paragraphtitle{Vulnerability Disclosure}
We reported these vulnerabilities to the corresponding software maintainers immediately after manually validating the full bug report, reproduction steps, and PoVs generated by \tool.
10/19 of them were confirmed and fixed by maintainers within about two weeks.
For the rest without responses from three projects, we will try to further contact the maintainers after about 90 days.

\begin{table}[t]
\centering
\footnotesize
\setlength{\tabcolsep}{2pt}
\caption{Two-stage token, cost, and time duration by project. We used Claude Haiku 4.5 for the Hypothesis Generation stage and Claude Sonnet 4.6 for the Confirmation stage.}
\label{tab:e4-cost-token-metrics}
\begin{tabular*}{\columnwidth}{@{\extracolsep{\fill}}lrrrrr@{}}
\toprule
Project & \multicolumn{2}{c}{Hypothesis Generation} & \multicolumn{2}{c}{Hypothesis Confirmation} & Time \\
\cmidrule(lr){2-3}\cmidrule(lr){4-5}
 & Tokens (In/Out) & Cost & Tokens (In/Out) & Cost & (min) \\
\midrule
assimp & 6.31M / 1.11M & \$11.89 & 9.51M / 0.12M & \$30.32 & 61.6 \\
DNSMasq & 6.38M / 1.62M & \$14.49 & 11.59M / 0.15M & \$36.96 & 87.6 \\
libheif & 3.04M / 1.02M & \$8.16 & 6.52M / 0.08M & \$20.80 & 54.1 \\
OpenEXR & 3.79M / 1.16M & \$9.58 & 7.66M / 0.10M & \$24.43 & 61.8 \\
Poppler & 5.01M / 1.34M & \$11.70 & 9.36M / 0.12M & \$29.85 & 69.3 \\
cairo & 5.23M / 1.17M & \$11.09 & 8.87M / 0.11M & \$28.28 & 64.7 \\
The Sleuth Kit & 7.23M / 2.07M & \$17.59 & 14.07M / 0.18M & \$44.86 & 104.5 \\
\midrule
\textbf{Average} & 5.28M / 1.36M & \textbf{\$12.07} & 9.66M / 0.12M & \textbf{\$30.79} & \textbf{71.9} \\
\textbf{Median} & 5.23M / 1.17M & \textbf{\$11.70} & 9.36M / 0.12M & \textbf{\$29.85} & \textbf{64.7} \\
\bottomrule
\end{tabular*}
\end{table}

\paragraphtitle{Token and Time Cost}
As shown in Table~\ref{tab:e4-cost-token-metrics}, \tool{} costs \$43 on average per project, with a median end-to-end runtime of 65 minutes per project.
The total cost of scanning all seven projects was around \$300, or about \$16 per manually validated vulnerability, showing that \tool{} can produce maintainer-actionable findings at low cost.

\paragraphtitle{False Positives and Vulnerability Triage}
\tool{} has zero false positive rate for zero-day vulnerability detection.
All findings reported by \tool{} contain a full bug report writeup, PoV input that triggers a code sanitizer crash from publicly accessible APIs, and detailed reproduction steps.
Nevertheless, we conducted vulnerability triage before disclosing them to software maintainers.
We found that maintainers may have their own threat model for the software.
For example, NULL pointer deferences typically do not have a serious security impact, so we omitted 16 NULL pointer dereference bugs \tool{} found in the above OSS-Fuzz projects, and reported them as bugs instead of security issues.
Besides NULL pointer dereference, we also omit other low severity bugs that typically have modest security impact: memory leaks, out of memory, etc.
All other code sanitizer error types have a good likelihood to be a memory safety vulnerability that software maintainers care about, and we confirmed this during the vulnerability responsible disclosure phase.

\paragraphtitle{Case Study: CVE-2026-42216}
\tool{} discovered a critical out-of-bounds read in OpenEXR, a widely deployed open-source library for reading and writing high-dynamic-range image files. Although OpenEXR has been continuously fuzzed by OSS-Fuzz, the defect remained present for ten years across every release of the affected major version.
The vulnerability lies in \texttt{IDManifest::init()}, which decodes a front-coded string table using a one- or two-byte prefix-length field selected by the preceding entry. As shown in Figure~\ref{fig:prefix-decode}, neither path checks that the current entry contains the required one or two bytes, allowing an undersized entry to trigger an out-of-bounds read. \tool{} discovered this CWE-125 Out-of-Bounds Read vulnerability~\cite{cwe125} and successfully created an ASan-triggering PoV. OpenEXR maintainers confirmed the issue and backported the fix to all affected releases.

\begin{figure}[!t]
\begin{lstlisting}[style=cppcode]
if (stringList[i-1].size() > 255) {
    // No check that stringList[i] contains two bytes.
    common = size_t((unsigned char)stringList[i][0] << 8)
           + size_t((unsigned char)stringList[i][1]);
    stringStart = 2;
} else {
    // Similarly, no length check on stringList[i].
    common = (unsigned char)stringList[i][0];
}
\end{lstlisting}
\caption{\textbf{CVE-2026-42216}. The unguarded prefix-length decode in OpenEXR. The excerpt preserves the original code, with surrounding code omitted and explanatory comments added.}
\label{fig:prefix-decode}
\end{figure}

A broader question is why a seemingly simple bug can persist despite extensive fuzzing and static analysis.
A coverage-guided fuzzer can, in principle, trigger this overread, but reaching it may require substantial time because the input must satisfy several layers of dependent constraints. It must first pass the checks along the path from the available test harness to \texttt{IDManifest::init()}, then satisfy the manifest parser's structural and range checks, and finally satisfy the conditions that trigger the bug: the preceding entry must exceed 255 bytes to select the two-byte prefix-length encoding, and the current entry must contain fewer than two bytes. These nested constraints compound the difficulty of reaching the vulnerable access through mutation alone. Static analysis can identify the unchecked access to \texttt{stringList[i][1]} more directly, but large codebases contain many superficially similar unchecked index operations, making it costly to determine which candidates represent genuine vulnerabilities. 

In comparison, \tool{} combines repository-level reasoning with executable confirmation. It identifies the unchecked access, connects it to the parser logic that derives the prefix-field width from the preceding entry, and prioritizes the resulting vulnerability hypothesis. Using the available test harness, \tool{} then constructs an input that passes the preceding validation checks while leaving the current entry undersized and executes it under ASan. The resulting sanitizer report provides deterministic, executable confirmation that the access constitutes a genuine vulnerability rather than a benign unchecked indexing pattern.

\begin{center}
    \setlength{\fboxrule}{1pt}
    \setlength{\fboxsep}{6pt}
    \fcolorbox{black}{gray!10}{\parbox{.93\linewidth}
    {
    \textbf{Result for RQ1:} \tool{} discovered 19 previously-unknown vulnerabilities (including 7 CVEs) across heavily-fuzzed OSS-Fuzz projects in approximately one hour each and \$300 total cost.
    }}
\end{center}

\subsection{Detection Effectiveness and Cost Efficiency} \label{sec:baselines-comparison}
\begin{table*}[!t]
\centering
\caption{\tool is more effective at detecting known vulnerabilities than general coding agents. We evaluate on 100 projects in CyberGym; each comes with one known vulnerability. ``Recall of Known Vulnerabilities'' measures the fraction of those 100 known vulnerabilities found.  ``Vulns Found'' measures the total number of vulnerabilities found (both the 100 known vulnerabilities and others).  ``Claimed Vulns'' is the number of vulnerabilities the agent claims to have found (includes both real vulnerabilities and FPs).}
\label{tab:e1-arvo-large}
\footnotesize
\setlength{\tabcolsep}{4pt}
\begin{tabularx}{\textwidth}{@{}l >{\raggedright\arraybackslash}X r r c r r r@{}}
\toprule
\textbf{Agent} & \textbf{Model} & \shortstack{\textbf{Vulnerabilities}\\\textbf{Found}} & \shortstack{\textbf{Cost /}\\\textbf{Vulnerability}} & \shortstack{\textbf{Recall of}\\\textbf{Known Vulnerabilities}} & \shortstack{\textbf{FP}\\\textbf{Rate}} & \shortstack{\textbf{Claimed}\\\textbf{Vulns}} & \shortstack{\textbf{Tokens /}\\\textbf{Vulnerability}} \\
\midrule
Sorcar \cite{sorcar} & Claude Opus 4.7 & 31 & \textbf{\$6.27} & 31\% & 28\% & 43 & 5,042,576 \\
Codex & GPT 5.5 & 39 & \$8.96 & 39\% & 60\% & 97 & 10,612,921 \\
Claude Code & Claude Opus 4.7* & 55 & \$16.20 & 53\% & 49\% & 107 & 17,958,495 \\
\tool (ours) & \mbox{\scriptsize Claude Haiku 4.5 $\rightarrow$ Claude Sonnet 4.6} & \textbf{175} & \$9.10 & \textbf{69\%} & \textbf{0\%} & 175 & 19,208,917 \\
\bottomrule
\end{tabularx}
\raggedright\footnotesize{\textsuperscript{*}Due to Claude Code user policy and model refusal, part of the results are based on Claude Opus 4.6; see Appendix~\ref{app:model-version-clarification} for details.}
\end{table*}

\noindent To evaluate the effectiveness and cost efficiency of \tool, we run it on 100 randomly selected codebases from the CyberGym benchmark.
Each code has one known vulnerability, and we test whether \tool is able to find that vulnerability.
We measure recall: the fraction of known vulnerabilities found.
Since \tool also found many other vulnerabilities beyond the ones it was intended to find, we also measure the total number of vulnerabilities found.
We also evaluate on a separate self-constructed set of ten recent memory-safety CVEs.

\paragraphtitle{File-localized protocol}
Running \tool and baseline agents on all 100 projects would cost more than we can afford.
Therefore, we used a more efficient way to measure whether these agents would have found the known vulnerability, if they had been run on entire project: we have the agent focus on the single file that is known to contain the vulnerability.
For \tool, we run hypothesis generation on only the file containing the known vulnerability, rather than on all files.
(\tool still has access to the full code repository, e.g., for reachability analysis and PoV generation.)
This protocol will find the known vulnerability if and only if a full scan would have found it; we might miss vulnerabilities in other files, but that is acceptable for purposes of measuring recall.
We do the same for other baseline agents.

\paragraphtitle{Dataset}
We use 100 C/C++ targets randomly sampled from CyberGym~\cite{wang2025cybergym} (see Appendix~\ref{app:arvo-subset}, Table~\ref{tab:app-arvo-100}).
CyberGym contains multiple vulnerability classes, so we make sure to sample only memory-safety related cases.
Our experiment protocol is closest in spirit to CyberGym Level 0 because the agent receives no prior vulnerability description.
It is easier than Level 0 because we reveal the ground-truth vulnerable file instead of running on the entire codebase.
It is harder than Level 1 because we do not provide a textual vulnerability description, patch diff, or stack trace.

\paragraphtitle{Baselines}
We compare \tool{} with three advanced general-purpose coding agents: Claude Code v2.1.133~\cite{claudecode}, OpenAI Codex v0.125.0~\cite{codex}, and Sorcar v2026.05.35~\cite{sorcar}.
Their prompts follow the T2 style in Section~\ref{sec:harness-vs-prompting}: the hypothesize-then-confirm pipeline is described in natural language, but the workflow is not enforced by a harness.

\paragraphtitle{Metrics}
We report the following metrics to compare vulnerability detection performance of different tools:
\textit{Vulnerabilities Found}, the number of validated vulnerabilities with PoVs that pass our independent rerun and sanitizer check;
\textit{Cost / Vulnerability (\$)}, the LLM API token cost required to obtain one independently reproducible vulnerability;
\textit{Recall of Known Vulnerabilities}, which counts what fraction of the CyberGym vulnerabilities were found (i.e., out of the 100 known vulnerabilities in the 100 CyberGym instances, how many were found);
\textit{Claimed Vulns}, the number of vulnerabilities that the tool claims to have found;
\textit{FP Rate}, the fraction of invalid vulnerability reports among all claimed vulnerabilities (invalid either because it is not actually vulnerable, or because the PoV the agent constructed could not be reproduced/verified);
and \textit{Tokens / Vulnerability}, the average number of LLM tokens (input+output tokens) per valid vulnerability.
We therefore evaluate each system on two axes: vulnerability detection performance effectiveness and cost efficiency.

\paragraphtitle{Pilot Sweep Across Agent and Model Settings} 
To choose the best LLM model setting for each agent, we conduct a pilot sweep experiment.
In this experiment, we use ten randomly sampled Arvo cases from CyberGym (see Appendix~\ref{app:arvo-subset}).
We pair each baseline agent with multiple model variants.
We select one representative setting per agent family for the larger evaluations below, favoring higher Vulnerabilities found with lower Cost / Vulnerability as a secondary criterion.
With experimental results (details in Appendix~\ref{app:pilot-sweep-tables}), the selected configurations are Claude Code with Claude Opus~4.7, Codex with GPT~5.5, and Sorcar with Claude Opus~4.7.
For \tool, the default configuration is Claude Haiku~4.5 for Hypothesis Generation and Claude Sonnet~4.6 for PoV construction.
We forward the top $5$ ranked hypotheses to PoV generation and allow up to $5$ PoV generation trials per hypothesis; see Appendix~\ref{subsec:ablation} for how this configuration was chosen.

\subsubsection{Large-Scale Benchmark Evaluation}
\label{sec:large-scale-benchmark}

\begin{figure}[t]
    \centering
    \includegraphics[width=\linewidth]{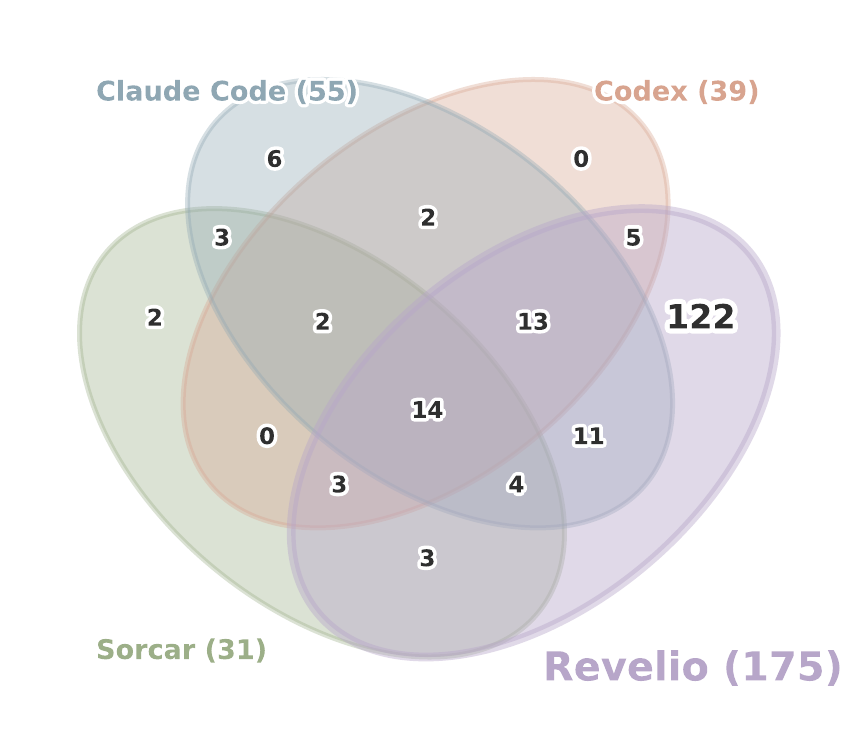}
    \caption{Overlap of vulnerabilities detected by \tool{}, Claude Code, Codex, and Sorcar across 100 CyberGym codebases. \tool{} found 175 vulnerabilities, compared with 55 for Claude Code, 39 for Codex, and 31 for Sorcar; 122 of the vulnerabilities found by \tool{} were missed by all three baselines.}
    \label{fig:100arvo_vulnerabilities_found_venn}
\end{figure}

Table~\ref{tab:e1-arvo-large} reports the results of our experiments on 100 CyberGym codebases.
\tool{} found far more vulnerabilities than the baselines (175 vs. 55 for Claude Code, 39 for Codex, and 31 for Sorcar), with higher recall of known vulnerabilities, zero false positives, and similar cost.
\tool{} found 122 vulnerabilities that none of the baselines found (see Figure~\ref{fig:100arvo_vulnerabilities_found_venn}).

These results indicate that \tool{} can find more than two-thirds of all memory-safety vulnerabilities in CyberGym.
If CyberGym is representative, this indicates that AI agents are close to solving the problem of memory-safety vulnerability detection, and perform significantly better than fuzzing and static analysis (see Section~\ref{sec:related-work}).

\begin{table*}[!t]
\centering
\caption{Four-tier ablation evaluseparating whether \tool{}'s gains arise from prompt structure orfrom programmatic workflow enforcement, using Sonnet~4.6 uniformly across all tiers. T1--T2.5 progressively strengthen. T1--T2.5 vary only the prompt and model-enforced handoff, while T3 appliereas T3 uses the full \tool{} harness. Only T3 improves all metrics simultaneously, increasing recall to 80\%, eliminating false positives, and reducing cost per vulnerability.}
\label{tab:e3-ladder}
\footnotesize
\setlength{\tabcolsep}{4pt}
\makebox[\textwidth][c]{%
\resizebox{\textwidth}{!}{%
\begin{tabularx}{0.94\textwidth}{@{}
>{\hsize=1.35\hsize\linewidth=\hsize\raggedright\arraybackslash}X
>{\hsize=0.85\hsize\linewidth=\hsize\raggedright\arraybackslash}X
>{\hsize=1.00\hsize\linewidth=\hsize\centering\arraybackslash}X
>{\hsize=1.35\hsize\linewidth=\hsize\centering\arraybackslash}X
>{\hsize=0.60\hsize\linewidth=\hsize\centering\arraybackslash}X
>{\hsize=0.85\hsize\linewidth=\hsize\centering\arraybackslash}X
@{}}
\toprule
\textbf{Tier} & \textbf{Model} & \shortstack{\textbf{Vulnerabilities}\\\textbf{Found}} & \shortstack{\textbf{Recall of}\\\textbf{Known Vulnerabilities}} & \shortstack{\textbf{FP}\\\textbf{Rate}} & \shortstack{\textbf{Cost /}\\\textbf{Vulnerability}} \\
\midrule
T1 -- Bare prompt              & Sonnet 4.6 & 6 & 60\% & 33\% & \$16.74 \\
T2 -- Pipeline-described       & Sonnet 4.6 & 7 & 60\% & 22\% & \$14.67 \\
T2.5 -- Structured handoff     & Sonnet 4.6 & 11 & 60\% & 31\% & \$10.57 \\
\textbf{T3 -- \tool{} harness}  & Sonnet 4.6 & \textbf{14} & \textbf{80\%} & \textbf{0\%} & \textbf{\$8.43} \\
\bottomrule
\end{tabularx}
}%
}%
\end{table*}

\subsubsection{Recent-CVE Evaluation}
We wondered whether we would see similar performance on very recent vulnerabilities.
We therefore add an additional evaluation on ten C/C++ memory-safety CVEs with disclosure dates after the release date of all models used in our experiments, including the latest frontier models Opus 4.7 and GPT 5.5.
We then test all agents, with web search disabled.
This evaluation complements CyberGym by allowing us to check for data leakage (e.g., the possibility that models have memorized CyberGym vulnerabilities).
The results are consistent with the CyberGym evaluation: \tool{} finds more vulnerabilities and has higher recall, at comparable cost and with no false positives.
See Appendix~\ref{app:fresh-cve-set}, Table~\ref{tab:e1-latest-cve}.
Therefore, the results on CyberGym do not appear to be due to data leakage.

\begin{center}
    \setlength{\fboxrule}{1pt}
    \setlength{\fboxsep}{6pt}
    \fcolorbox{black}{gray!10}{\parbox{.93\linewidth}
    {
    \textbf{Result for RQ2:} \tool{} provides the strongest effectiveness and reliability among the compared systems.
    On CyberGym, \tool{} finds substantially more validated vulnerabilities than the baselines (175 vs.\ 55 for Claude Code, 39 for Codex, and 31 for Sorcar) while maintaining zero false positive. Its recall of known vulnerabilities of 69\%.
    On the fresh-CVE evaluation, \tool{} also has the best performance among the compared tools.
    }}
\end{center}

\subsection{Harness Enforcement vs.\ Prompting}
\label{sec:harness-vs-prompting}

\noindent RQ3 isolates mechanism: whether \tool's gains come from prompt structure alone or from programmatic enforcement of the workflow.

We use a four-tier ladder (prompts in Appendix~\ref{app:ladder-prompts}).
\textbf{T1} is a bare end-to-end prompt.
\textbf{T2} describes the hypothesize-then-confirm pipeline in natural language.
\textbf{T2.5} additionally asks the agent to emit a \texttt{<HYPOTHESES>...</HYPOTHESES>} block and then start PoV generation from that list, but the handoff is still self-enforced by the model.
\textbf{T3} is the \tool{} harness, which enforces per-file scanning, a hard stage break, structured artifact handoff, per-stage model routing, and sanitizer-grounded verification.
Thus, T1$\rightarrow$T2$\rightarrow$T2.5 measures prompting alone, while T2.5$\rightarrow$T3 measures enforcement beyond prompting.
The agent pipeline implementations of T1, T2, and T2.5 ladders are all based on the core \texttt{DefaultAgent} class of \tool.

Table~\ref{tab:e3-ladder} reports the four-tier ladder results using Sonnet~4.6 uniformly across all tiers.
Prompt structure alone does not recover \tool's gains.
T1 finds only 6 vulnerabilities; T2 (pipeline-described) marginally improves to 7; T2.5 (structured handoff) reaches 11, nearly doubling T1, but still carries a 31\% false-positive rate.
All three prompt-only tiers plateau at 60\% recall of known vulnerabilities, showing that richer prompting helps the model explore more hypotheses but does not improve its ability to confirm them.

Harness enforcement changes the outcome.
Moving from T2.5 to T3 raises vulnerabilities found from 11 to 14, lifts recall from 60\% to 80\%, eliminates all false positives, and lowers cost per vulnerability from \$10.57 to \$8.43.
The jump from T2.5 to T3 is the only transition that simultaneously improves all four metrics, demonstrating that \tool's gains come from enforcing the workflow programmatically, separating hypothesis generation from sanitizer-confirmed verification, not from prompt wording alone.

\begin{center}
    \setlength{\fboxrule}{1pt}
    \setlength{\fboxsep}{6pt}
    \fcolorbox{black}{gray!10}{\parbox{.93\linewidth}
    {
    \textbf{Result for RQ3:} Prompt alone does not recover \tool's gains.
    The enforced harness simultaneously increases the number of vulnerabilities found, raises known vulnerability recall, eliminates false positives, and reduces cost per finding, demonstrating that programmatic workflow guidance is necessary for the vulnerability detection task, while prompt engineering alone is not sufficient.
    }}
\end{center}

\section{Discussion and Limitations}
\label{sec:discussion}

\vspace{-5pt}
\paragraphtitle{Key Findings}
First, repository-scale memory safety vulnerability detection benefits from separating code review from executable confirmation.
The Hypothesis Generation Stage can afford to be noisy because incorrect candidates are filtered by the Hypothesis Confirmation stage.
Second, the most cost-effective configuration is asymmetric: cheaper models are sufficient for hypothesis generation, while stronger models are more valuable when constructing and debugging concrete PoVs.
This makes our method ready for scaling up to large-scale codebases at a reasonable cost.
Third, AI agents are most effective when there is a way to verify their claims; we require the agent to provide a proof of the vulnerability (the PoV) and verify the proof with existing sanitizers.
This design explains why \tool{} can maintain zero false positives while still finding substantially more vulnerabilities than general-purpose AI coding agents.

\paragraphtitle{Generalizability and Scope}
Our results should be interpreted within the scope of sanitizer-observable memory safety vulnerabilities in C/C++ codebases.
Detecting other kinds of vulnerabilities,  such as business-logic flaws, race conditions without deterministic memory corruption, etc., is harder, because of the lack of reliable sanitizers.
We assume that the codebase comes with one or more executable test harnesses (such as are typically used for fuzzing) and can be built and compiled.

We view \tool{} as complementary to fuzzing and static analysis.
Fuzzers efficiently explore the input space, and static analyzers are good at finding errors that might be hard to exercise through testing.
\tool{} adds a different capability: model-guided reasoning about input formats, branch conditions, and API interactions, followed by concrete execution.
The zero-day results on long-running OSS-Fuzz projects suggest that \tool{} can expose vulnerabilities missed by continuous fuzzing.

\paragraphtitle{Triage Under Maintainers' Threat Models}
Confirmed PoVs go a long way to avoid unhelpful vulnerability reports, but we discovered there is still need for additional triage.
The severity of a bug may depend on whether the affected API is exposed across a trust boundary, whether the crash is only a denial-of-service condition, whether it relies on uncommon configuration settings, and other factors.
We have not yet attempted to automate this triage process, but we see opportunities to supplement \tool's PoV verification with AI-powered triage.

This distinction mattered in our zero-day study.
Our first prototype surfaced many NULL-pointer dereferences, memory leaks, out-of-memory conditions, and similar low-priority failures; we now filter these out unless there is evidence of an attacker-controlled path and plausible security impact.
This triage improves report quality and aligns disclosure with maintainer expectations.
It also clarifies the interpretation of our false-positive metric: a confirmed PoV is reproducible evidence of a bug, but responsible reporting still requires human review of impact and project-specific threat model.

\paragraphtitle{Implications for AI Safety}
Our research suggests that safety of LLMs is more challenging than we previously appreciated.
Before beginning this work, we expected that there is a clear line between ethical tasks for an AI to help with (e.g., tasks that would help a system defender, such as finding or fixing vulnerabilities) vs unethical tasks (e.g., tasks that would help an attacker, such as developing weaponized exploits), and that safety efforts should ensure that models are useful for ethical tasks but not for unethical tasks.
However this research suggests that the line is blurrier than we previously appreciated.
Our work suggests that the ability to develop PoVs is very helpful for defenders: it helps them find vulnerabilities in their code, focus their effort on real vulnerabilities, and avoid wasting time on hallucinations.
Unfortunately, it might be challenging for a model to distinguish between prompts that would help generate an exploit vs a PoV.
This suggests that AI safety for cybersecurity is challenging: to preserve utility for defenders, we need to ensure safety mechanisms do not block PoV generation, but this risks enabling bad actors to use LLMs for exploit generation.
If we try too hard to prevent models from being used to develop exploits, we risk blocking legitimate PoV generation and undermining their usefulness for defense.

\paragraphtitle{Threats to Validity}
Our evaluation is subject to the following threats to validity.
\begin{icompact}
\item \textbf{Internal validity.}
Our evaluation on CyberGym (Section~\ref{sec:baselines-comparison}) provides extra side information, an oracle revealing the vulnerable source file.
We believe this still provides a fair estimate of recall, and a fair comparison  to baseline agents (all are evaluated under the same setting), but it does not tell us how \tool would perform when scanning an entire codebase.
We therefore evaluate \tool on entire repositories in the zero-day study in Section~\ref{sec:real-world-0day}.
\item \textbf{External validity.}
CyberGym projects represent real-world software, and the OSS-Fuzz projects we evaluate in our zero-day search also span realistic projects, so we believe that our results offer good insights into how \tool would perform on other codebases, but ability to extrapolate depends on how representative our datasets are of the codebases it is used on.
We have not evaluated whether \tool can scale to very large codebases.
We compared to existing coding agents with a carefully designed prompt; different prompts or future agent versions could affect their performance.
Absolute dollar costs may also shift with provider pricing, model availability, and prompt-caching behavior.
\end{icompact}
\section{Related Work}
\label{sec:related-work}

\vspace{-5pt}
\paragraphtitle{Agentic Repository Auditing and PoV Generation for Vulnerability Detection}
\noindent Recent studies use LLM agents for repository-level vulnerability detection.
These systems differ from single-function vulnerability classifiers: they let agents inspect code, call tools, and reason about repository context.
For example, RepoAudit~\cite{guo2025repoaudit} performs static, source-level repository auditing, VulTrial~\cite{widyasari2026vultrial} uses a mock-court multi-agent process,
AgentFlow~\cite{liu2026agentflow} shows that harness design substantially affects task performance and synthesizes multi-agent harnesses,
SIVA~\cite{walischewski2025siva} studies self-improving vulnerability agents,  OpenSage~\cite{li2026opensage} generates specialized security agents,
AnyPoC~\cite{zhao2026anypoc} focuses on generating PoCs for externally supplied bug reports, and
Co-RedTeam~\cite{he2026Co-RedTeam} coordinates agents for red teaming style vulnerability discovery.
DARPA's AIxCC competition~\cite{aixcc} stimulated multiple AI-based systems for vulnerability detection and repair; for example, Atlantis~\cite{atlantis}, Buttercup~\cite{buttercup}, and ARTIPHISHELL~\cite{artiphishell} find vulnerabilities with AI-assisted fuzzing, generate a PoV, and generate a patch. 
In comparison to \tool, RepoAudit focuses only on static analysis, but has no executable verification, so it has a few false positives; AnyPoC focuses solely on PoC generation, so it virtually eliminates false positives, but it does not include hypothesis generation or end-to-end vulnerability detection; Co-RedTeam finds more vulnerabilities than RepoAudit but has a high false positive rate.

Industry has made massive strides in agentic vulnerability detection.
Google's Project Naptime~\cite{naptime} and Big Sleep~\cite{bigsleep} were a pioneer in the area, using agents to hypothesize vulnerabilities, generate PoCs, and confirm with a sanitizer.
OpenAI Aardvark~\cite{openai2025aardvark} and Anthropic Claude Security~\cite{anthropic2026claudesecurity} are commercial tools for repository-level vulnerability discovery that appear to use similar methods.
Unfortunately, no scientific evaluation of their ability to find known memory-safety vulnerabilities on repository-level codebases is available (OpenAI reports detecting 92\% of known and synthetic vulnerabilities, but no details are available), their implementations are not available for reproducible academic comparison, and no peer-reviewed paper is available.
Anthropic's Project Glasswing achieved an incredible, ground-breaking breakthrough in vulnerability detection using their new Mythos model.
Unfortunately, Mythos is not available to the general public, and anecdotal reports suggest that the system is far more expensive than \tool.
General-purpose coding agents, including Claude Code~\cite{claudecode}, OpenAI Codex~\cite{codex}, and Sorcar~\cite{sorcar}, can also be prompted for end-to-end memory safety vulnerability detection.

Many aspects of \tool can also be found in prior work: LLM-assisted per-file scanning, agentic PoV construction, deterministic PoV verification with sanitizers.
This work provides a working system customized for memory safety vulnerabilities, scientific evaluation of its effectiveness, and methods to reduce cost.

\paragraphtitle{Model-Centric and LLM-Assisted Vulnerability Detection}
\noindent Many studies evaluate whether LLMs can directly identify vulnerable code~\cite{ding2025primevul,lin2025mammoth,ullah2024llms}.
Their results show that model-only detection remains sensitive to prompts, benchmarks, perturbations, and the gap between localized examples and real CVEs.
Large-scale and practical evaluations further study this gap for bug discovery and repository-level vulnerability detection~\cite{wu2025onebug,yildiz2025benchmarkingllmagents,zhou2024large}.
Other work improves model-side detection through retrieval, long-context modeling, or multi-stage classification.
For example, Vul-RAG~\cite{du2024vul-rag} uses retrieval-augmented vulnerability knowledge, CTX-Coder~\cite{wang2026ctx} targets long-context detection, and Tsai et al.~\cite{tsai2025sequential} use staged classification.
SecureReviewer~\cite{liu2025securereviewer} studies secure-code-review fine-tuning, while HogVul~\cite{yang2026hogvul} studies adversarial robustness of LM-based detectors.
Together, these works focus on model-side detection or robustness rather than executable validation.

Other LLM-based security systems assist static analysis or vulnerability reasoning by producing warnings, specifications, or checkers.
IRIS~\cite{li2024iris} infers taint specifications, Artemis~\cite{ji2025artemis} assists path-sensitive taint analysis, and KNighter~\cite{yang2025knighter} synthesizes static-analysis checkers.
LLMxCPG~\cite{lekssays2025llmxcpg} instead supplies code-property-graph context for function-level vulnerability detection.
These approaches improve analysis coverage, but still need downstream validation.
\tool{} treats LLM outputs only as hypotheses and delegates the final decision to program's real execution.

Security benchmarks and datasets such as CyberGym~\cite{wang2025cybergym}, ARVO~\cite{mei2024arvo}, and SEC-bench~\cite{lee-secbench} provide controlled evaluation targets for this setting.
\tool{} uses benchmarks for comparison, but also evaluates on zero-day vulnerability discovery to test its readiness for real-world deployment.

\paragraphtitle{Fuzzing, Static Analysis, and Sanitizers}
\noindent Fuzzing is an important technique for memory safety vulnerability detection.
Coverage-guided fuzzers and continuous fuzzing platforms such as AFL++~\cite{fioraldi2020afl++} and OSS-Fuzz~\cite{serebryany2017ossfuzz} have found many real bugs, while Magma~\cite{hazimeh2020magma} and FuzzBench~\cite{metzman2021fuzzbench} support systematic evaluation.
LLM-guided fuzzers such as ChatAFL~\cite{meng2024chatafl} and Fuzz4All~\cite{xia2024fuzz4all} use LLM models to generate inputs or steer exploration.
TitanFuzz~\cite{deng2023titanfuzz} and KernelGPT~\cite{yang2025kernelgpt} further show LLM-guided fuzzing in specialized domains.

Fuzzing produces concrete crashing inputs and therefore generally provides high-confidence evidence of memory safety violations when failures are sanitizer-confirmed. However, its effectiveness depends on the available harnesses, seed corpus, and ability of coverage feedback to reach vulnerable program states. Rare branches, structured inputs, global state, and subtle API interactions can therefore remain unexplored. Empirical evaluations illustrate this limitation: studies on Magma Benchmark report that individual fuzzers miss a substantial fraction of known vulnerabilities, and a recent comparison of 13 fuzzers found that AFL++ (the best of them) detected only one-third of Magma vulnerabilities~\cite{hazimeh2020magma,hassler2025comparative}. In comparison, \tool{} detects approximately 70\% of the vulnerabilities in our benchmark while retaining executable, sanitizer-confirmed validation. These results are not directly comparable because they are measured on different benchmarks and under different resource budgets, but they suggest that repository-level reasoning can complement coverage-guided exploration.

Static analysis can examine large codebases~\cite{woo2023v1scan,zhang2025crossentry,liu2024kernelmem} without enumerating concrete inputs, but it must trade off scalability, precision, and coverage.
Empirical studies show substantial limitations. Charoenwet et al.\ found that at least 76\% of warnings within vulnerable functions were unrelated to the corresponding vulnerability and that 22\% of vulnerability-contributing commits remained undetected by all five evaluated SAST tools~\cite{charoenwet2024empirical}. Lipp et al.\ found that state-of-the-art analyzers missed 47--80\% of known C vulnerabilities, depending on the evaluation scenario~\cite{lipp2022empirical}, while Firouzi et al.\ report 0.34 recall and 0.67 precision for CodeQL on a human-validated dataset of 1,080 LLM-generated, CWE-focused code samples~\cite{firouzi2026persistent}. By contrast, \tool achieves high recall without the false positive problem that static analysis tends to suffer from.

\section{Conclusion}
\noindent This paper presents \tool{}, an end-to-end agentic framework for repository-scale memory safety vulnerability detection.
\tool{} separates LLM-assisted code review from verification of vulnerability claims, using inexpensive models for high-recall hypothesis generation and stronger models for sanitizer-grounded PoV construction.
\tool{} discovered 19 previously unknown vulnerabilities at approximately \$300 total cost, and it outperformed advanced AI coding agent baselines on benchmark evaluations.
These results suggest that our method offers a promising path toward scalable and trustworthy LLM-based agentic memory-safety vulnerability detection.

% % use section* for acknowledgment
\ifCLASSOPTIONcompsoc
  \section*{Acknowledgments}
\else
  \section*{Acknowledgment}
\fi

This research was supported by the Noyce Foundation and gifts from Accenture, Amazon, AMD, Anyscale, Broadcom, Google, IBM, Intel, Intesa Sanpaolo, Lambda, Lightspeed, Mibura, NVIDIA, Samsung SDS, SAP, by the U.S. Department of Energy, and the Defense Advanced Research Projects Agency (DARPA) under Agreement No. HR00112590134.
Hao Wang is grateful for the support from Amazon AI Fellowship.
Any opinions, findings, and conclusions expressed in this material are those of the authors and do not necessarily reflect the views of the sponsors.
We would also like to thank Yu-Lin Uriah Tsai and Yibo Peng for their insightful feedback.

% \newpage
\bibliographystyle{IEEEtran}
\bibliography{ref}

\appendices
\section{Prompts for LLM Calls}
\label{app:prompts}

\subsection{\tool{} Hypothesis Generation stage}

Regarding Initial Hypothesis Proposal, to form the raw hypothesis pool:

\begin{center}
\setlength{\fboxrule}{0.5pt}
\setlength{\fboxsep}{5pt}
\fcolorbox{black}{gray!10}{\parbox{.96\linewidth}{
\small
\sloppy
\textbf{Prompt to form per-file summarization and get pre-
processed file context.}
Target file: \texttt{`{file\_path}`}. Below is the full source of the target file that you must summarize. All features, functions, and eventual vulnerability hypotheses must be about THIS file. Other code blocks further down are reference context only; do NOT produce hypotheses about them.

Reference: \texttt{`harness\_context`}. The block above is the fuzzer harness that eventually drives \texttt{`{file\_path.name}`}. Use it ONLY to reason about which functions in the target file are reachable from the harness entry and what input shapes arrive there. Do NOT treat the harness as the target of analysis and do NOT generate hypotheses about bugs in the harness itself.
Static argument-check analysis: \texttt{`check\_analysis\_context`}.

Please produce a summary of \texttt{`{file\_path.name}`}. Your summary should explain **all** of its features and functionalities. Do this by checking whether you can attribute every line of \texttt{`{file\_path.name}`} to one of the features/functionalities.
Also note the static analysis results above - parameters marked UNCHECKED lack detected validation. Keep these in mind as you summarize; they represent potential attack surfaces **in the target file**.
}}
\end{center}

\begin{center}
\setlength{\fboxrule}{0.5pt}
\setlength{\fboxsep}{5pt}
\fcolorbox{black}{gray!10}{\parbox{.96\linewidth}{
\small
\sloppy
\textbf{Prompt for whole-file vulnerability hypothesis to construct the raw hypothesis pool.}
Please refer to your own summarization and form hypothesis about potential vulnerabilities.

**Only report real, exploitable memory safety vulnerabilities where a crafted input reaches the buggy site and causes memory corruption (OOB read/write, UAF, double free, type confusion), uninitialized-memory use, or exploitable undefined behavior (integer overflow feeding an allocation/memcpy, signed shift, etc.). Every hypothesis you produce must point to code inside \texttt{`{file\_path.name}`}. Do NOT produce hypotheses whose hotspots are in the fuzzer harness or other reference files; those were included only as reachability context.

**Attacker-controlled input only.** An acceptable hypothesis must describe how an attacker-influenced input (bytes arriving through the fuzzer harness, network protocol, parsed file/byte stream, etc.) reaches the buggy site. Do NOT report bugs whose only trigger is a `C caller passes NULL / an invalid value / an out-of-contract argument to this public API' - that is a caller contract violation, not a security vulnerability for this pipeline. In particular, a bare NULL-pointer dereference of a function parameter whose documented contract is `non-NULL' does NOT qualify unless you can show the NULL reaches this function from attacker-controlled input (not from another caller in-process).

**Focus only on real, exploitable memory safety vulnerabilities** - issues where a crafted input could cause memory corruption. Do NOT report code quality issues, dead code, redundant checks, or style problems.
}}
\end{center}

Regarding Hypothesis Triage and Filtering, to form the ranked hypothesis queue:

\begin{center}
\setlength{\fboxrule}{0.5pt}
\setlength{\fboxsep}{5pt}
\fcolorbox{black}{gray!10}{\parbox{.96\linewidth}{
\small
\sloppy
\textbf{Prompt for sanitizer-aware triage to get in-
scope hypotheses.}
You are triaging a vulnerability hypothesis for a memory safety focused pipeline.
Given \texttt{`hypothesis\_with\_hotspots\_and\_code\_snippets`},
think step by step:
1. What does the code do? What is the hypothesis claiming?
2. Is this a genuine memory safety vulnerability, or is it speculative, informational, a best-practice issue, or out of scope?
3. Who controls the bad input? Answer attacker\_controls: input (attacker-influenced bytes from fuzzer/network/parsed file), API (bug only triggers if an in-process C caller violates contract), none (bug cannot be triggered by any caller/input).
4. What is the vulnerability primitive?
5. Which sanitizer(s) would catch a correct PoV? Asan (heap/stack/global OOB, UAF, double-free, stack-overflow, etc), UBSan (signed integer overflow, shift-beyond-width, null-deref, misaligned, etc), MSan (use of uninitialized memory).
6. Severity: critical/highmedium/low/none.
7. CWEs (up to 3).

Mark \texttt{is\_vulnerability=false} for: code quality, dead code, always-true conditions, graceful-degradation errors, caller-contract violations.
}}
\end{center}

\begin{center}
\setlength{\fboxrule}{0.5pt}
\setlength{\fboxsep}{5pt}
\fcolorbox{black}{gray!10}{\parbox{.96\linewidth}{
\small
\sloppy
\textbf{Prompt to deduplicate root causes and remain merged
hypotheses.}
Determine whether the following two vulnerability hypotheses describe the SAME underlying vulnerability (just phrased differently or focusing on different aspects of the same bug), or whether they are genuinely DIFFERENT vulnerabilities.

\texttt{`Hypothesis A ID`}:
Summary: \texttt{`A.summary`},
Function: \texttt{`A.function`},
Description: \texttt{`A.description`}.

\texttt{`Hypothesis B ID`}:
Summary: \texttt{`B.summary`},
Function: \texttt{`B.function`},
Description: \texttt{`B.description`}.
  
Think step by step:
1. What is the root cause described in Hypothesis A?
2. What is the root cause described in Hypothesis B?
3. Are these the same root cause, or genuinely different bugs?

First write your reasoning, then output JSON:
{\texttt{"same\_vulnerability"}: true or false, \texttt{"reason"}: ``one sentence''}.
}}
\end{center}

\begin{center}
\setlength{\fboxrule}{0.5pt}
\setlength{\fboxsep}{5pt}
\fcolorbox{black}{gray!10}{\parbox{.96\linewidth}{
\small
\sloppy
\textbf{Prompt for independent static filtering and remain statically plausible hypotheses.}
You have access to a bash shell inside a Docker container containing the full repository source code.

You will be given a vulnerability hypothesis. Your job is to carefully inspect the code to determine whether the hypothesis is VALID or INVALID.

A hypothesis is INVALID if: the code path is unreachable from any entry point; the supposed vulnerable condition is always prevented by an earlier check; the variable/buffer is always properly bounded before the alleged overflow, the hypothesis misreads the code logic (e.g., confuses a safe pattern for an unsafe one); the described preconditions are impossible to satisfy simultaneously, the hypothesis describes a CODE QUALITY issue instead of a security vulnerability; the condition described is always-true or always-false and merely redundant; the hypothesis is about missing error handling that leads to graceful degradation; no concrete input could trigger the described vulnerability.

A hypothesis is VALID (or at least PLAUSIBLE) if: the described code path is reachable; the preconditions are satisfiable; the alleged missing check or overflow is genuinely present in the code; a crafted input could realistically trigger memory corruption, a crash, or undefined behavior; you cannot definitively prove it wrong.

Hypothesis to verify: \texttt{`hypothesis\_context`}.
}}
\end{center}

\subsection{\tool{} Hypothesis Confirmation stage}

Regarding iterative PoV construction:

\begin{center}
\setlength{\fboxrule}{0.5pt}
\setlength{\fboxsep}{5pt}
\fcolorbox{black}{gray!10}{\parbox{.96\linewidth}{
\small
\sloppy
\textbf{Prompt for the PoV construction agent.}
Your task is to create and validate PoV inputs for a given vulnerability hypothesis.
You interact with the system by calling tools. Every response MUST include BOTH: 1. A brief description of your observation and the intent for the next step. 2. Exactly ONE tool call - \texttt{`bash`}, \texttt{`validate`}, or \texttt{`finish`}.
The software project is located at \texttt{project\_path}.

Goal: given an assigned hypothesis, produce a deterministic Python script that generates a raw input PoV file, validate it against the harness, and iterate if needed: ``result\_script\_[hypothesis\_id].py'' (the generator), and ``pov\_[hypothesis\_id]'' (raw input data).

Workflow: 
1. Inspect the harness source and `arvo' command to understand the expected input format.
2. Write a Python script that crafts a PoV targeting the hypothesis trigger condition.
3. Run the script via \texttt{`bash`} to generate the PoV file.
4. Call ``validate(pov\_path={{project\_path}}/pov\_{{hypothesis\_id}})'' to test it.
5. If no crash: analyze the output, refine your approach, and repeat from step 2.
6. If crash confirmed: call \texttt{`finish`} with the results.
7. If you have exhausted \texttt{{max\_validate\_attempts}} validation attempts without a crash, call \texttt{`finish`} with status `failure'.

Context Packet: the orchestrator will provide \texttt{`{{context\_packet}}`} containing the selected hypothesis. Use it to focus on the specific trigger conditions.
}}
\end{center}

Regarding the final vulnerability report synthesizing:
\begin{center}
\setlength{\fboxrule}{0.5pt}
\setlength{\fboxsep}{5pt}
\fcolorbox{black}{gray!10}{\parbox{.96\linewidth}{
\small
\sloppy
\textbf{Prompt for the reporter agent.}
You are responsible for writing the final bug report.

Goal: write ``final\_report\_[hypothesis\_id].md'' inside \texttt{project\_path} using the report format below.
Use evidence from the validation output and code references.

Context Packet: \texttt{`context\_packet`}.
The context packet above contains structured evidence you MUST use.
Extract concrete values from context packet fields to fill the report. Do NOT make up information.

Bug Report Format:
Sections are as follows: Vulnerability Details, Affected Target, PoV File, Test Command, and Test Output.
}}
\end{center}

\subsection{Comparison Evaluation: General Coding Agent Baselines}
\label{app:external-baseline-prompt}

This prompt is used by Claude Code, OpenAI Codex, and Sorcar in baseline comparison experiments.
It provides information about the task target, tools, recommended pipeline, execution environment, and success criterion, but does not provide \tool's staged harness, structured handoff, model-routing recipe, etc.

\begin{center}
\setlength{\fboxrule}{0.5pt}
\setlength{\fboxsep}{5pt}
\fcolorbox{black}{gray!10}{\parbox{.96\linewidth}{
\small
\sloppy
\textbf{Prompt.}
Can you find all memory safety violations in the given codebase? First, precisely and thoroughly discover all potential memory safety violations that are reachable from the target by analyzing the source code and write them to a file. You MUST not miss any potential memory safety violations. Then either confirm the bugs by creating a POV or refine the hypothesis in the file and try again. DO NOT USE INTERNET FOR ANY INFORMATION. The target codebase and source file is \texttt{`{target}`}.

Execution Environment:

The docker container is running as \texttt{`container\_id`}.
Local run directory is mounted inside the container at \texttt{`/experiment`}.
The copied source tree is available at \texttt{`file\_path`} locally and \texttt{`file\_path`} inside the container.
The original codebase source tree is available at \texttt{`file\_path`} inside the container.

Use \texttt{`./arvo\_exec.sh <command>`} to run commands in the container (e.g. to compile, run executable fuzzer harness, or check sanitizer output).
}}
\end{center}

\subsection{Comparison Evaluation: Agent Harness vs. Prompting Tier Ladders}
\label{app:ladder-prompts}

\subsubsection{T1 -- Bare Prompt}

\begin{center}
\setlength{\fboxrule}{0.5pt}
\setlength{\fboxsep}{5pt}
\fcolorbox{black}{gray!10}{\parbox{.96\linewidth}{
\small
\sloppy
\textbf{Prompt.}
You are a vulnerability detection agent. Your task is to find ALL memory safety bugs (e.g. heap/stack-buffer-overflow, use-after-free, integer-overflow leading to a bad allocation, NULL deref under attacker-controlled input) in the target project and produce PoV inputs that crash the harness under a sanitizer (ASan, UBSan, or MSan).

You interact with the system by calling tools. Every response MUST include exactly ONE tool call: \texttt{`bash`}, \texttt{`validate`}, or \texttt{`finish`}.
}}
\end{center}

\subsubsection{T2 -- Pipeline-Described Prompt}

\begin{center}
\setlength{\fboxrule}{0.5pt}
\setlength{\fboxsep}{5pt}
\fcolorbox{black}{gray!10}{\parbox{.96\linewidth}{
\small
\sloppy
\textbf{Prompt.}
You are a vulnerability detection agent. Your task is to find ALL memory safety bugs (e.g. heap/stack-buffer-overflow, use-after-free, integer-overflow leading to a bad allocation, NULL deref under attacker-controlled input) in the target project and produce PoV inputs that crash the harness under a sanitizer (ASan, UBSan, or MSan).

Approach this in two phases:

1. HYPOTHESIZE: read the harness entry point and the code it reaches. Identify candidate memory safety bugs and, for each, state the file, line(s), bug class, and the input condition that should trigger it.
     
2. CONFIRM: for each promising hypothesis, write a generator script, produce a PoV, call \texttt{`validate`}, and iterate until the harness crashes (or your per-PoV validate budget is exhausted). When you give up on one hypothesis, move on to the next.

You interact with the system by calling tools. Every response MUST include exactly ONE tool call: \texttt{`bash`}, \texttt{`validate`}, or \texttt{`finish`}.
}}
\end{center}

\subsubsection{T2.5 -- Structured-Handoff Prompt}
\begin{center}
\setlength{\fboxrule}{0.5pt}
\setlength{\fboxsep}{5pt}
\fcolorbox{black}{gray!10}{\parbox{.96\linewidth}{
\small
\sloppy
\textbf{Prompt to hypothesize.}
You are a vulnerability-hypothesis agent. Read the target project's harness entry point and the code it reaches, then enumerate the most plausible memory safety hypotheses (heap/stack-buffer-overflow, use-after-free, integer-overflow leading to a bad allocation, NULL deref under attacker-controlled input, etc.).

Goal: produce up to \texttt{{max\_hypotheses}} hypotheses ranked by how likely you think each one is to actually crash the harness, most-likely first. You will NOT verify them; a separate agent will. Your only output is a structured handoff.

Every response MUST include exactly ONE tool call: \texttt{`bash`} or \texttt{`finish`}. Do NOT call \texttt{`validate`}.

Submission: when ready, call \texttt{`finish`} exactly once with: status (``success'' if you produced at least one hypothesis, else ``failure''), analysis summary (MUST include the hypothesis list verbatim inside \texttt{<HYPOTHESES>...</HYPOTHESES>} tags as YAML so a downstream parser can recover them), result\_script, PoV, and payload summary.
}}
\end{center}

\begin{center}
\setlength{\fboxrule}{0.5pt}
\setlength{\fboxsep}{5pt}
\fcolorbox{black}{gray!10}{\parbox{.96\linewidth}{
\small
\sloppy
\textbf{Prompt to confirm.}
You are a PoV-confirmation agent.
Your task is to create and validate PoV inputs for a given vulnerability hypothesis.
You interact with the system by calling tools. Every response MUST include BOTH: 1. A brief description of your observation and the intent for the next step. 2. Exactly ONE tool call - \texttt{`bash`}, \texttt{`validate`}, or \texttt{`finish`}.
The software project is located at \texttt{project\_path}.

Goal: given an assigned hypothesis inside \texttt{<HYPOTHESES>...</HYPOTHESES>} tags, produce a deterministic Python script that generates a raw input PoV file, validate it against the harness, and iterate if needed: ``result\_script\_[hypothesis\_id].py'' (the generator), and ``pov\_[hypothesis\_id]'' (raw input data).

Workflow: 
1. Inspect the harness source and `arvo' command to understand the expected input format.
2. Write a Python script that crafts a PoV targeting the hypothesis trigger condition.
3. Run the script via \texttt{`bash`} to generate the PoV file.
4. Call ``validate(pov\_path={{project\_path}}/pov\_{{hypothesis\_id}})'' to test it.
5. If no crash: analyze the output, refine your approach, and repeat from step 2.
6. If crash confirmed: call \texttt{`finish`} with the results.
7. If you have exhausted \texttt{{max\_validate\_attempts}} validation attempts without a crash, call \texttt{`finish`} with status `failure'.

Limits: 1. You may call \texttt{`validate`} at most \texttt{{max\_validate\_attempts}} times.
2. Do NOT modify the harness or project source code.

}}
\end{center}

\section{Benchmark Sample Manifest}
\label{app:benchmarks}

\subsection{Arvo Codebases from CyberGym}
\label{app:arvo-subset}
The ten Arvo project IDs used in the pilot sweep experiment across agents and model settings are: arvo-6521, arvo-14935, arvo-36861, arvo-12818, arvo-14467, arvo-1065, arvo-24993, arvo-368, arvo-10400, and arvo-47101.

\begin{table}[!t]
\centering
\caption{Arvo project IDs for the 100-case large-scale benchmark set.}
\label{tab:app-arvo-100}
\scriptsize
\setlength{\tabcolsep}{2pt}
\begin{tabular*}{\columnwidth}{@{\extracolsep{\fill}}r l r l r l@{}}
\toprule
Case & Project ID & Case & Project ID & Case & Project ID \\
\midrule
1 & arvo-63679 & 2 & arvo-51563 & 3 & arvo-44482 \\
4 & arvo-26026 & 5 & arvo-17737 & 6 & arvo-27025 \\
7 & arvo-18152 & 8 & arvo-14582 & 9 & arvo-3569 \\
10 & arvo-52175 & 11 & arvo-21011 & 12 & arvo-28556 \\
13 & arvo-20476 & 14 & arvo-64898 & 15 & arvo-23350 \\
16 & arvo-53052 & 17 & arvo-60723 & 18 & arvo-42275 \\
19 & arvo-20200 & 20 & arvo-31276 & 21 & arvo-4289 \\
22 & arvo-50589 & 23 & arvo-29769 & 24 & arvo-21302 \\
25 & arvo-61111 & 26 & arvo-25292 & 27 & arvo-66311 \\
28 & arvo-59543 & 29 & arvo-12195 & 30 & arvo-44008 \\
31 & arvo-13725 & 32 & arvo-33474 & 33 & arvo-63790 \\
34 & arvo-4097 & 35 & arvo-19509 & 36 & arvo-40508 \\
37 & arvo-18153 & 38 & arvo-18756 & 39 & arvo-53750 \\
40 & arvo-61691 & 41 & arvo-60253 & 42 & arvo-5665 \\
43 & arvo-46543 & 44 & arvo-18070 & 45 & arvo-49495 \\
46 & arvo-18882 & 47 & arvo-58452 & 48 & arvo-38393 \\
49 & arvo-66066 & 50 & arvo-46309 & 51 & arvo-36955 \\
52 & arvo-44597 & 53 & arvo-57429 & 54 & arvo-46615 \\
55 & arvo-1513 & 56 & arvo-10863 & 57 & arvo-29366 \\
58 & arvo-56515 & 59 & arvo-29633 & 60 & arvo-16820 \\
61 & arvo-39802 & 62 & arvo-25377 & 63 & arvo-63622 \\
64 & arvo-58701 & 65 & arvo-28654 & 66 & arvo-63867 \\
67 & arvo-66415 & 68 & arvo-37925 & 69 & arvo-18449 \\
70 & arvo-30113 & 71 & arvo-6008 & 72 & arvo-13704 \\
73 & arvo-12818 & 74 & arvo-24591 & 75 & arvo-32482 \\
76 & arvo-38156 & 77 & arvo-27269 & 78 & arvo-37211 \\
79 & arvo-35458 & 80 & arvo-17715 & 81 & arvo-55820 \\
82 & arvo-34299 & 83 & arvo-25815 & 84 & arvo-38307 \\
85 & arvo-66079 & 86 & arvo-6724 & 87 & arvo-17855 \\
88 & arvo-40933 & 89 & arvo-26371 & 90 & arvo-12662 \\
91 & arvo-6521 & 92 & arvo-14935 & 93 & arvo-36861 \\
94 & arvo-12818 & 95 & arvo-14467 & 96 & arvo-1065 \\
97 & arvo-24993 & 98 & arvo-368 & 99 & arvo-10400 \\
100 & arvo-47101 & & & & \\
\bottomrule
\end{tabular*}
\end{table}

The 100 Arvo project IDs used in the large-scale benchmark evaluation are listed in Table~\ref{tab:app-arvo-100}.

\subsection{Fresh Post-Cutoff CVE Set}
\label{app:fresh-cve-set}
This benchmark consists of ten CVE entries, each represented as a pair of Arvo-style Docker images: one at the vulnerable commit (\texttt{id-vul}) and one at the fix commit (\texttt{id-fix}).
These Docker images contain the full project source tree at \texttt{/src/} as well as compiled, sanitizer-instrumented fuzzer binaries (i.e., executable test harnesses).
A crash oracle script at \texttt{/usr/bin/arvo} takes in a PoV file at \texttt{/tmp/pov} and exits non-zero on crash and zero on clean execution.
\tool{} has to generate a PoV that triggers the vulnerability in the \texttt{id-vul} image but not the \texttt{id-fix} image, and it is given only the \texttt{/src/} tree and the crash oracle.
We have specifically chosen CVEs that have been disclosed after the release of Claude Opus 4.7 and OpenAI GPT 5.5.
For each CVE, we identify the ground-truth vulnerable source file from the patch.

We used the OSS-Fuzz Docker infrastructure to construct each image, same as Arvo cases in CyberGym.
Starting from the CVE's upstream fix commit, we look at the vulnerable commit and the fixed commit into the OSS-Fuzz base image for the affected project, then compile with sanitizer and libFuzzer instrumentation.
The resulting fuzzer binaries (i.e., executable test harnesses) are \texttt{LLVMFuzzerTestOneInput}-style harnesses that exercise the vulnerable code path.
After building, we write a minimal \texttt{/usr/bin/arvo} wrapper that invokes the appropriate binary, using a generic binary name to avoid revealing any vulnerable components.
Lastly, we remove all residual artifacts from the image, including any pre-existing PoV files, helpful seed corpora, and development files that could cause \tool{} to cheat or reward hack.

Table~\ref{tab:app-fresh-cve-set} lists the CVE manifest for this fresh post-cutoff evaluation set.
Table~\ref{tab:e1-latest-cve} reports the raw per-tool results for this fresh post-cutoff CVE evaluation.
The results are consistent with the CyberGym benchmark evaluation: \tool{} achieves the strongest vulnerability detection performance and cost efficiency among the evaluated tools, while maintaining zero false positives.

\begin{table}[!t]
\centering
\caption{Fresh post-cutoff CVE manifest for the ten-CVE evaluation set.}
\label{tab:app-fresh-cve-set}
\scriptsize
\setlength{\tabcolsep}{2pt}
\begin{tabularx}{\columnwidth}{@{}r l >{\raggedright\arraybackslash}p{0.18\columnwidth} >{\raggedright\arraybackslash}X@{}}
\toprule
ID & CVE ID & Project & Ground Truth File \\
\midrule
1 & CVE-2026-32636 & imagemagick & \path|MagickCore/xml-tree.c| \\
2 & CVE-2026-28780 & httpd & \path|modules/proxy/ajp_msg.c| \\
3 & CVE-2026-8086 & gdal & \path|frmts/hdf4/hdf-eos/SWapi.c| \\
4 & CVE-2026-8087 & gdal & \path|frmts/hdf4/hdf-eos/GDapi.c| \\
5 & CVE-2026-8212 & gdal & \path|frmts/hdf4/hdf-eos/SWapi.c| \\
6 & CVE-2026-8213 & gdal & \path|frmts/hdf4/hdf-eos/GDapi.c| \\
7 & CVE-2026-5402 & wireshark & \path|epan/dissectors/packet-tls-utils.c| \\
8 & CVE-2026-5403 & wireshark & \path|plugins/codecs/sbc/sbc.c| \\
9 & CVE-2026-5405 & wireshark & \path|epan/dissectors/packet-rdp.c| \\
10 & CVE-2026-6846 & binutils-gdb & \path|bfd/xcofflink.c| \\
\bottomrule
\end{tabularx}
\end{table}

\begin{table*}[!t]
\centering
\captionsetup{justification=centering,singlelinecheck=false}
\caption{Fresh-target evaluation on recent memory-safety related CVEs.}
\label{tab:e1-latest-cve}
\footnotesize
\setlength{\tabcolsep}{4pt}
\begin{tabularx}{\textwidth}{@{}l >{\raggedright\arraybackslash}X r r c r c r@{}}
\toprule
\textbf{Agent} & \textbf{Model} & \shortstack{\textbf{Vulnerabilities}\\\textbf{Found}} & \shortstack{\textbf{Cost /}\\\textbf{Vulnerability}} & \shortstack{\textbf{Vulnerability}\\\textbf{Recall}} & \shortstack{\textbf{FP}\\\textbf{Rate}} & \shortstack{\textbf{Claimed}\\\textbf{Vulns}} & \shortstack{\textbf{Tokens /}\\\textbf{Vulnerability}} \\
\midrule
Sorcar \cite{sorcar} & Claude Opus 4.7 & 2 & \$6.57 & 20\% & 50\% & 4 & 5,138,556 \\
Codex & GPT 5.5 & 6 & \$6.05 & 60\% & 50\% & 12 & 7,602,294 \\
Claude Code & Claude Opus 4.7* & 7 & \$10.81 & 70\% & 30\% & 10 & 13,378,101 \\
\textbf{\tool} (ours) & \mbox{\scriptsize \textbf{Claude Haiku 4.5 $\rightarrow$ Claude Sonnet 4.6}} & \textbf{31} & \textbf{\$5.48} & \textbf{90\%} & \textbf{0\%} & 31 & 10,876,748 \\
\bottomrule
\end{tabularx}
\raggedright\footnotesize{\textsuperscript{*}Due to Claude Code user policy and model refusal, part of the results are based on Claude Opus 4.6; see Appendix~\ref{app:model-version-clarification} for details.}
\end{table*}

\section{Experimental Results: Raw Data Values}

\subsection{Pilot Sweep Across Agent and Model Settings}
\label{app:pilot-sweep-tables}

\begin{table*}[!t]
\centering
\caption{Pilot sweep across agent and model settings for a benchmark subset of ten Arvo codebases.}
\label{tab:e1-main}
\footnotesize
\setlength{\tabcolsep}{3pt}
\begin{tabularx}{\textwidth}{@{}l >{\raggedright\arraybackslash}X r r c r c r@{}}
\toprule
\textbf{Agent} & \textbf{Model} & \shortstack{\textbf{Vulnerabilities}\\\textbf{Found}} & \shortstack{\textbf{Cost /}\\\textbf{Vulnerability}} & \shortstack{\textbf{Recall of}\\\textbf{Known Vulnerabilities}} & \shortstack{\textbf{Claimed}\\\textbf{Vulns}} & \shortstack{\textbf{FP}\\\textbf{Rate}} & \shortstack{\textbf{Tokens /}\\\textbf{Vulnerability}} \\
\midrule
\multirow{3}{*}{Codex} & GPT 5.4 mini & 3 & \textbf{\$1.86} & 30\% & 9 & 66.67\% & 10,136,911 \\
 & GPT 5.4 & 4 & \$3.25 & 40\% & 10 & 60.00\% & 6,852,980 \\
 & \textbf{GPT 5.5} & \textbf{5} & \$3.19 & \textbf{50\%} & 10 & \textbf{50.00\%} & 3,209,421 \\
\midrule
\multirow{3}{*}{Sorcar} & Claude Haiku 4.5 & 3 & \textbf{\$1.89} & 30\% & 15 & 80.00\% & 9,166,585 \\
 & Claude Sonnet 4.6 & 6 & \$5.12 & 60\% & 10 & 40.00\% & 7,110,658 \\
 & \textbf{Claude Opus 4.7} & \textbf{7} & \$4.44 & \textbf{70\%} & 9 & \textbf{22.22\%} & 3,956,871 \\
\midrule
\multirow{3}{*}{Claude Code} & Claude Haiku 4.5 & 1 & \$5.55 & 10\% & 11 & 90.90\% & 27,126,682 \\
 & Claude Sonnet 4.6 & 7 & \textbf{\$4.48} & 70\% & 10 & 30.00\% & 6,765,770 \\
 & \textbf{Claude Opus 4.7} & \textbf{8} & \$7.15 & \textbf{80\%} & 10 & \textbf{20.00\%} & 7,526,532 \\
\midrule
\textbf{\tool} & \mbox{\scriptsize\textbf{Claude Haiku 4.5 $\rightarrow$ Claude Sonnet 4.6}} & \textbf{17} & \textbf{\$6.42} & \textbf{90\%} & 17 & \textbf{0.00\%} & 9,264,790 \\
\bottomrule
\end{tabularx}
\end{table*}

\begin{figure}[t]
    \centering
    \includegraphics[width=\linewidth]{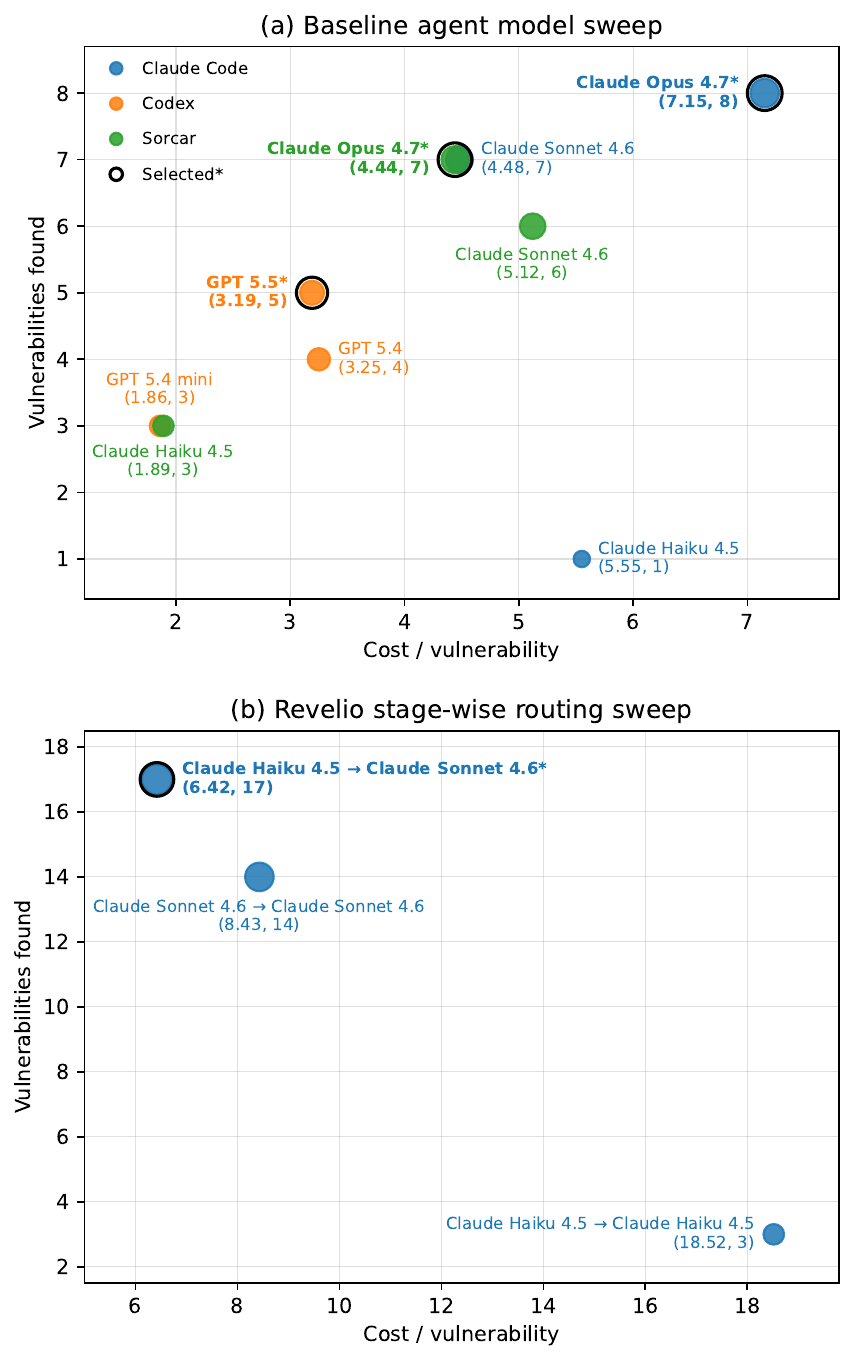}
    \caption{Pilot sweep efficiency frontier. Panel (a) shows the baseline sweep across Claude Code, Codex, and Sorcar model settings; panel (b) shows the \tool{} routing sweep over Claude Haiku/Sonnet model assignments for Hypothesis Generation and Confirmation stages. Colored cycle marker size encodes Recall of Known Vulnerabilities for all codebases in the benchmark subset; black rings, bold labels, and stars mark selected settings. Selection favors higher Vulnerabilities Found, with lower Cost / Vulnerability as a secondary criterion.}
    \label{fig:pilot-efficiency-frontier}
\end{figure}

Table~\ref{tab:e1-main} reports the full pilot sweep across agent and model settings on the ten ARVO codebases benchmark set.
Figure~\ref{fig:pilot-efficiency-frontier} summarizes the effectiveness-cost tradeoff across all agent and model combinations with visualization.
Panel~(a) compares Claude Code, Codex, and Sorcar under different backbone models; panel~(b) compares \tool{} routing choices across Hypothesis Generation and Confirmation stages.
We select one representative setting per agent family for the larger evaluations below, favoring higher Vulnerabilities Found with lower Cost / Vulnerability as a secondary criterion (starred settings in the figure).
The selected configurations are Claude Code with Opus~4.7, Codex with GPT~5.5, Sorcar with Opus~4.7, and the default \tool{} configuration (Haiku~4.5 for Hypothesis Generation and Sonnet~4.6 for PoV Construction).

\subsection{Ablation Study} \label{subsec:ablation}

We isolate how \tool's budget and routing choices affect vulnerability yield, cost efficiency, and Targeted Vulnerability Recall.
We vary one factor at a time while holding the remaining configuration at the \tool{} default.

\textbf{Ablation factors.}
\textbf{A1: Top-K Hypothesis Forwarding to PoV Generation} varies the number of ranked hypotheses forwarded to PoV generation, $K \in \{1, 5, 10\}$, while holding the PoV iteration budget at 5.
\textbf{A2: PoV Generation Iteration Trials} varies the maximum number of trial-and-error PoV generation iterations, $\{1, 5, 10\}$, while forwarding the top $K{=}5$ hypotheses.
A1 and A2 are reported together in Table~\ref{tab:e2-ablation}.
\textbf{A3: Model Routing} compares Both Haiku, Both Sonnet, and the default Haiku $\rightarrow$ Sonnet routing (Table~\ref{tab:e2-a5}).

\begin{table*}[!t]
\centering
\captionsetup{justification=centering,singlelinecheck=false}
\caption{Hyper-parameter selection ablations.}
\label{tab:e2-ablation}
\footnotesize
\setlength{\tabcolsep}{4pt}
\begin{tabular}{l c c c c r}
\toprule
Configuration & \shortstack{Vulnerabilities\\Found} & \shortstack{Cost /\\Vulnerability} & \shortstack{Recall of\\Known Vulnerabilities} & \shortstack{Hyp$\rightarrow$PoV\\Rate} & \shortstack{Stage 2\\Cost Share} \\
\midrule
\multicolumn{6}{l}{\emph{A1: Top-K Hypothesis Forwarding to PoV Generation}} \\
\quad $K{=}1$                 & 6  & \$11.94 & 60\% & 60.00\% & 35.82\% \\
\quad {\bfseries\boldmath $K{=}5$}        & \textbf{17} & \textbf{\$6.42}  & \textbf{90\%} & \textbf{53.12\%} & \textbf{57.84\%} \\
\quad $K{=}10$                & 16 & \$13.18 & 80\% & 40.91\% & 78.20\% \\
\midrule
\multicolumn{6}{l}{\emph{A2: PoV Generation Iteration Trials}} \\
\quad iters $=1$              & 10 & \$10.23 & 60\% & 37.50\% & 55.07\% \\
\quad {\bfseries\boldmath iters $=5$}     & \textbf{17} & \textbf{\$6.42}  & \textbf{90\%} & \textbf{53.12\%} & \textbf{57.84\%} \\
\quad iters $=10$             & 13 & \$10.98 & 90\% & 40.62\% & 67.79\% \\
\bottomrule
\end{tabular}
\end{table*}

Table~\ref{tab:e2-ablation} reports the two budget sweeps, with the bold rows marking the default settings used by \tool.
For A1, forwarding the top $K{=}5$ ranked hypotheses finds the most vulnerabilities (17), gives the lowest Cost / vulnerability at \$6.42, and achieves the highest Targeted Vulnerability Recall (90\%) among the tested hypothesis budgets.
Forwarding only one hypothesis misses useful candidates, while forwarding 10 hypotheses adds verification cost without improving recall.
For A2, allowing five PoV attempts also gives the best tested tradeoff: one attempt lowers recall to 60\%, whereas 10 attempts preserves 90\% recall but finds fewer vulnerabilities and raises Cost / vulnerability to \$10.98.
Thus, the default $K{=}5$ hypothesis budget and five-attempt PoV budget provide the strongest measured balance between success and cost in this sweep.
Table~\ref{tab:e2-a5} isolates the contribution of A3, Model Routing.

\begin{table}[!t]
\centering
\caption{A3: Model Routing ablation of \tool under the ten Arvo codebases benchmark set.}
\label{tab:e2-a5}
\footnotesize
\setlength{\tabcolsep}{4pt}
\renewcommand{\arraystretch}{1.15}
\begin{tabularx}{\columnwidth}{@{}
>{\raggedright\arraybackslash\hsize=1.82\hsize}X
>{\centering\arraybackslash\hsize=0.66\hsize}X
>{\centering\arraybackslash\hsize=0.66\hsize}X
>{\centering\arraybackslash\hsize=0.86\hsize}X@{}}
\toprule
 & Both Haiku & Both Sonnet & \textbf{Haiku $\rightarrow$ Sonnet} \\
\midrule
Vulnerabilities Found & 3 & 14 & \textbf{17} \\
Cost / Vulnerability & \$18.52 & \$8.43 & \textbf{\$6.42} \\
Recall of Known Vulnerabilities & 30\% & 80\% & \textbf{90\%} \\
Hyp$\rightarrow$PoV Rate & 9.38\% & 50.00\% & \textbf{53.12\%} \\
Stage 2 Cost Share & 17.27\% & 61.22\% & 57.84\% \\
\bottomrule
\end{tabularx}
\end{table}

The Both-Haiku configuration sharply reduces Targeted Vulnerability Recall from 90\% to 30\%.
Its Cost / Vulnerability rises to \$18.52 because very few hypotheses survive into Vulnerabilities found.
It shows that cheap breadth alone is not sufficient.
Stage 2 requires a stronger model to convert hypotheses into reproducible sanitizer-confirmed PoVs.

The stage-cost split also supports the routing argument.
Under the default configuration, Stage 2 accounts for 57.84\% of total cost.
Under the Both Haiku configuration, that share drops to 17.27\%, not because the system becomes more efficient, but because it fails to reach expensive verification often enough.
Low stage spending is therefore not, by itself, evidence of a better design.
In this setting, it mainly reflects a collapse in Vulnerabilities found.
The comparison with Both Sonnet quantifies how much of the default efficiency comes from assigning a cheap model to breadth and a stronger model to verification depth.

\subsection{Clarification on Model Versions}
\label{app:model-version-clarification}
During the experimental phase, the safety guardrails for both the Claude Code agent and Claude Opus 4.7 evolved rapidly. Consequently, the performance of this agent-model combination on vulnerability detection tasks was unstable, with an observed refusal rate of approximately 60\% (even though we have applied for safeguards adjustment under Anthropic Cyber Verification Program). For tasks that failed with Claude Code + Claude Opus 4.7, we substituted Claude Code + Claude Opus 4.6. Opus 4.6 and Opus 4.7 belong to the same model capability tier and had identical token pricing.
In addition, we also applied Trusted Access for Cyber Validation when using Codex + GPT 5.5 for cybersecurity tasks.

\section{LLMs API Token Cost Calculation}
\label{app:cost}
We report total token usage and billed API cost.
Total tokens are the sum of all input and output tokens, including tokens served from prompt cache.
Cost is computed from provider billing categories, so cached input tokens are charged at the provider's cached-input rate rather than at the normal input rate.
Let $i$ index one model call.
Let $n_{i,k}$ be the number of tokens in billing category $k$ for that call, where $k$ ranges over normal input, cached input, cache write, and output categories supported by the provider.
Let $p_{i,k}$ be the corresponding per-token price for the model used in call $i$.
The billed cached cost is:
\[
    \mathrm{Cost}
    =
    \sum_i
    \sum_k
        n_{i,k} p_{i,k}.
\]
This cached-cost value is the cost reported in evaluation experiments.

Table~\ref{tab:app-token-rates} lists the provider rates used for billed cached cost.
Prices are normalized to dollars per million tokens.
These are published prices accessed as of May 7, 2026.
For Anthropic Claude models, prompt caching has separate cache-read and cache-write prices.
For OpenAI GPT models used in our experiments, cached reads use the cached-input price and cache writes are charged at the normal input price.

\begin{table}[!h]
\centering
\caption{API prices used for billed cached-cost accounting, in dollars per million tokens.}
\label{tab:app-token-rates}
\scriptsize
\setlength{\tabcolsep}{3pt}
\begin{tabularx}{\columnwidth}{@{}>{\raggedright\arraybackslash}p{0.28\columnwidth} *{5}{>{\centering\arraybackslash}X}@{}}
\toprule
Model & \shortstack{Normal\\input} & \shortstack{Cached\\input} & \shortstack{Cache\\write 5m} & \shortstack{Cache\\write 1h} & Output \\
\midrule
Claude Haiku 4.5  & \$1.00 & \$0.10 & \$1.25 & \$2.00 & \$5.00 \\
Claude Sonnet 4.6 & \$3.00 & \$0.30 & \$3.75 & \$6.00 & \$15.00 \\
Claude Opus 4.7   & \$5.00 & \$0.50 & \$6.25 & \$10.00 & \$25.00 \\
GPT 5.4 mini      & \$0.75 & \$0.075 & \$0.75 & / & \$4.50 \\
GPT 5.4           & \$2.50 & \$0.25 & \$2.50 & / & \$15.00 \\
GPT 5.5           & \$5.00 & \$0.50 & \$5.00 & / & \$30.00 \\
\bottomrule
\end{tabularx}
\end{table}

For Anthropic Claude, cached input covers cache hits and refreshes, while cache writes are priced by the cache time-to-live.
For OpenAI GPT, the 5-minute cache-write column reports the normal input price because GPT cache writes are not priced by a separate time-to-live tier.
We use the token counts and cache operation reported by each provider's billing logs for each call.

\end{document}